\documentclass[aps,prb,twocolumn,floatfix,longbibliography]{revtex4-1}

\usepackage{graphicx}
\usepackage{float}
\usepackage{amsfonts,amsmath,amssymb}
\usepackage{amsthm}
\usepackage[normalem]{ulem}
\usepackage{dsfont,bm}
\usepackage{amsbsy}
\usepackage[colorlinks=true,linkcolor=blue,pagecolor=blue,filecolor=blue,menucolor=blue,urlcolor=blue,citecolor=blue,anchorcolor=blue]{hyperref}
\usepackage{mathbbol}
\usepackage{appendix}
\usepackage{sidecap}
\usepackage{xcolor}

\begin{document}

\title{Parity-driven RKKY decoupling and anomalous $1/R$ Dzyaloshinskii-Moriya interaction in odd-parity magnets}

\author{Morteza Salehi}
\email{m.salehi@basu.ac.ir}
\affiliation{Department of Physics, Bu-Ali Sina University, Hamadan, Iran}

\author{Tohid Farajollahpour}

\affiliation{Department of Physics, Norwegian University of Science and Technology (NTNU), NO-7491 Trondheim, Norway}

\date{\today}

\begin{abstract}
Unconventional odd-parity magnets, characterized by an odd-parity momentum-dependent spin splitting, offer a fundamentally distinct paradigm for non-collinear spintronics. Here, we theoretically investigate the Ruderman-Kittel-Kasuya-Yosida indirect exchange in two-dimensional odd-parity magnets subjected to Rashba spin-orbit coupling. Using an analytical real-space Green's function formalism, we uncover a parity-driven spatial decoupling in the magnetic response. Because of the odd-parity exchange field, the out-of-plane Ising interaction is structurally insulated from the macroscopic magnetic modulation, oscillating isotropically at the shifted Fermi wavevector. Conversely, the in-plane Heisenberg components exhibit pronounced, directionally tunable spatial beating. Beyond collinear exchange, the hybridized bands generate a highly tunable, three-component Dzyaloshinskii-Moriya interaction alongside symmetric off-diagonal anisotropies. We reveal that the out-of-plane chiral twisting is driven by the massive, nonrelativistic momentum shift, while the in-plane chiral components are relativistic. Furthermore, the competition between the nodal geometry and the Rashba gap drives an anomalous, dimension-reducing crossover, in which the in-plane chiral components follow a 1D-like $1/R$ spatial decay along the nodal lines over an extended intermediate-distance window before ultimately recovering the conventional 2D $1/R^2$ asymptote. We demonstrate that these findings hold universally across the odd-parity class by explicitly generalizing our framework to $f$-wave magnets. Finally, numerical tight-binding calculations verify the striking robustness of this parity-driven decoupling against both scalar and random magnetic exchange disorder. These findings establish odd-parity magnets as promising platforms for engineering robust, directionally tunable non-collinear spin textures.
\end{abstract}
\maketitle

\section{Introduction}
Magnetic materials have traditionally been divided into two major classes, namely ferromagnets, which carry a macroscopic net magnetization, and antiferromagnets, in which compensated antiparallel sublattices cancel it\cite{Marder2010Book,Tsymbal2019Book,Blundell2001Book}. This long-standing dichotomy was recently overturned by the discovery of unconventional magnets, most notably altermagnets, which combine the vanishing net magnetization of antiferromagnets with a strong, momentum-dependent spin splitting normally associated with ferromagnets \cite{Smejkal2022, Smejkal_review2022, Mazin2022,Jungwirth2025Newton,Mazin2021PNAS,Mazin2023PRB}. Enforced by crystalline spin-group symmetries rather than relativistic spin-orbit coupling, this splitting is even in momentum and carries $d$-, $g$-, or $i$-wave character, and it already underpins a growing set of spintronic functionalities, including nonrelativistic spin-polarized currents and giant tunneling magnetoresistance \cite{SmejkalMR2022,Herasymchuk2025PRB,Bode2007,Das2023JPCM}, a spontaneous crystal Hall response \cite{Smejkal2020,Reichlova2024NC}, and efficient electrical spin-current generation \cite{GonzalezHernandez2021}, with the predicted spin splitting now confirmed in experiments \cite{Krempasky2024,Bai2022PRL,Fedchenko2024SciAdv,Ding2024PRL,Lee2024PRL,Osumi2024PRB,Orlova2024JETP}.

Following this breakthrough, an even more distinctive class of magnetic materials was proposed, the odd-parity magnets, including $p$-wave and $f$-wave systems \cite{Hellenes2024,Brekke2024,Jungwirth2024,Jungwirth2025Newton,Song2025Nature, Ezawa2025PRB}. Stabilized by a combination of time-reversal and half-translation symmetries, these magnets host an odd-parity, non-relativistic momentum-dependent spin splitting \cite{Brekke2024} that manifests as a rigid, spin-dependent shift of the Fermi surfaces in momentum space, leaving the carrier effective mass unchanged. They have been predicted to host highly efficient non-relativistic Edelstein effects \cite{Chakraborty2024}, to support unconventional transverse spin currents at metallic junctions \cite{Hedayati2025,Yuan2026FOP}, to enable purely electrical detection of the N\'{e}el vector through linear and nonlinear conductivities \cite{Ezawa2024,Maeda2024JPSJ}, and to give rise to exotic superconducting proximity effects and topological phases \cite{Sukhachov_SC2025}.

A particularly sensitive probe of such band structures is supplied by magnetic impurities, which couple to one another indirectly through the surrounding conduction electrons via the Ruderman-Kittel-Kasuya-Yosida (RKKY) interaction \cite{Ruderman1954, Kasuya1956, Yosida1957,Wang2017PRB}. Because the RKKY range function is governed by the static spin susceptibility of the host, it faithfully encodes the Fermi-surface geometry and the underlying spin symmetries, and it has consequently served as a diagnostic of indirect exchange in a broad range of low-dimensional \cite{Yafet1987PRB,Imamura2004,Chesi2010PRB,Lyu2007JAP,Kernreiter2013PRB} and Dirac materials \cite{Chang2015PRB}, including graphene \cite{Kogan2011PRB,Saremi2007, BlackSchaffer2010, Sherafati2011} and the surfaces of topological insulators \cite{Liu2009, Zhu2011,Shiranzaei2017PRB,Zyuzin2014PRB}, and it has been measured directly both in layered metallic structures \cite{ParkinMauri1991} and between individual magnetic adatoms \cite{Zhou2010}. In systems lacking inversion symmetry, Rashba spin-orbit coupling (RSOC) locks the electron spin to its momentum \cite{Bychkov1984, Manchon2015,Bihlmayer2015NJP,Soumyanaryyanan2016Nature} and twists the indirect exchange, generating an antisymmetric, non-collinear coupling between the impurities known as the Dzyaloshinskii-Moriya (DM) interaction \cite{Dzyaloshinskii1958, Moriya1960,Yang2015PRL}. This RKKY-mediated chiral coupling was established by Fert and Levy \cite{FertLevy1980}, and in Rashba electron gases it was shown to take the form of a twisted exchange interaction \cite{Imamura2004} and to generate a DM coupling at surfaces \cite{Crepieux1998}. In spin-orbit-coupled hosts the indirect exchange generically decomposes into competing Heisenberg, Ising, and DM contributions \cite{Chesi2010PRB, Zhu2011}. The interplay of these channels is a cornerstone for engineering twisted, non-collinear spin textures such as spin spirals and magnetic skyrmions \cite{Bode2007, Muhlbauer2009, Heinze2011, Nagaosa2013,Thiaville2012EL} for next-generation memory and logic devices \cite{FertCros2013, FertReyrenCros2017, Romming2013, Parkin2008}.
\begin{figure}[t]
    \centering
    \includegraphics[width=0.5\textwidth]{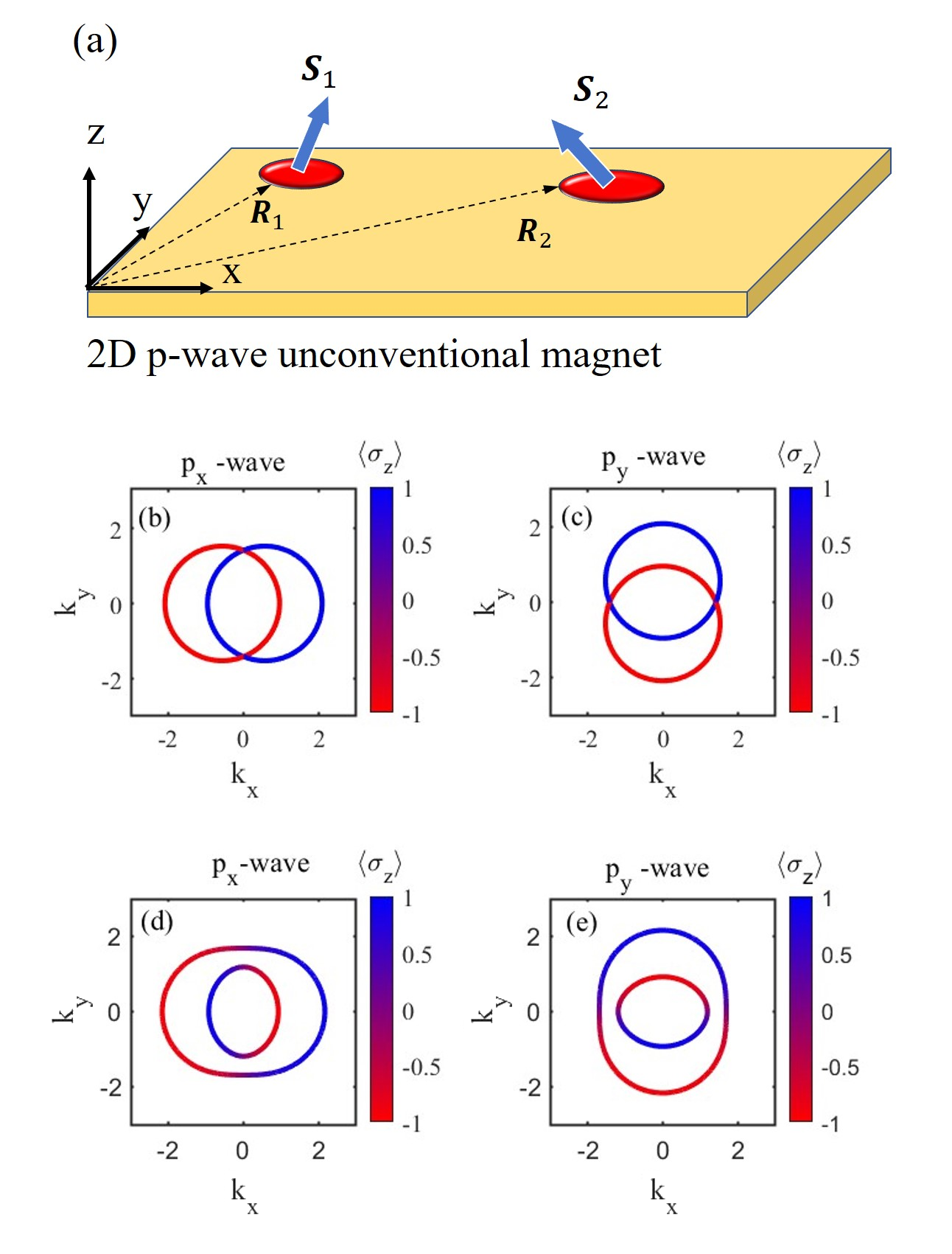}
    \caption{ (a) Schematic illustration of 2D $p$-wave unconventional magnet in the presence of two magnetic impurities with spins $\bm{S}_1$ and $\bm{S}_2$, located at spatial positions $\bm{R}_1$ and $\bm{R}_2$. The localized impurity spins interact with each other via the indirect RKKY interaction mediated by the itinerant electrons. (b) to (e) Constant-energy contours in momentum space showing the spin-splitting effect. The color map corresponds to the out-of-plane spin expectation value $\langle\sigma_z\rangle$, highlighting the momentum-dependent spin polarization. The top row shows the unperturbed (b) $p_x$-wave ($\beta = 0$) and (c) $p_y$-wave ($\beta = \pi/2$) unconventional magnets in the absence of RSOC, ($\lambda = 0$), where the bands exhibit a uniform momentum shift with perfectly intersecting nodes. The bottom row displays the corresponding (d) $p_x$-wave and (e) $p_y$-wave magnets in the presence of RSOC (dimensionless Rashba splitting $\lambda k_F/\mathcal{E}_F = 1/4$). The introduction of RSOC lifts the nodal degeneracies, inducing band hybridization and a smooth mixing of the spin projections driven by the competition between the $p$-wave magnetic order and relativistic spin-momentum locking. }
    \label{fig:rashba_interaction}
\end{figure}

To date, RKKY studies of unconventional magnets have addressed only the even-parity case, predicting directional beating and tunable anisotropic DM oscillations in $d$-wave altermagnets \cite{Amundsen2024, Yarmohammadi2025,Yarmohammadi2026}, while the $p$-wave case remains unexplored. The closest related result is that impurity-induced Friedel oscillations, the one-body counterpart of RKKY, exhibit distinct spatial decoupling and beating in $p$-wave magnets \cite{Sukhachov_Friedel2024}.

In this work, we theoretically investigate the RKKY interaction between two magnetic impurities embedded in a two-dimensional odd-parity magnet subjected to RSOC, as schematically depicted in Fig.~\ref{fig:rashba_interaction}(a). Using an analytical long-distance real-space Green's-function approach, we demonstrate that the odd-parity symmetry of the magnetic order fundamentally alters the RKKY exchange tensor compared to the $d$-wave case\cite{Yarmohammadi2025,Amundsen2024}. Because the magnetic order shifts the Fermi surfaces, we uncover a profound decoupling in the spatial magnetic responses. Specifically, we find that the out-of-plane Ising interaction does not exhibit macroscopic beating and oscillates at the shifted Fermi-surface radial wave vector. In contrast, the in-plane Heisenberg components exhibit spatial beating. We show that the chiral and symmetric anisotropic exchange terms are governed by a hierarchy of energy scales yielding distinct spatial regimes. Most notably, near the nodal directions, the retention of the RSOC-induced gap drives an anomalous, dimension-reducing $1/R$ spatial scaling. To demonstrate the universality of these phenomena, we focus our primary derivations on $p$-wave magnets and generalize our analytical framework to $f$-wave magnets. Finally, we employ a real-space tight-binding model to verify that this parity-driven decoupling is highly robust against both scalar impurities and random magnetic exchange disorder. Ultimately, our findings establish that the magnetic landscape in odd-parity magnets is qualitatively distinct from ferromagnets and even-parity altermagnets, offering new theoretical insights for non-collinear spintronics.

\section{THEORY}
\label{sec:theory}

\subsection{Model Hamiltonian for 2D $p$-wave Magnets}
We characterize the electronic properties of a two-dimensional (2D) unconventional $p$-wave magnet in the presence of RSOC by utilizing an effective two-band continuum Hamiltonian. In momentum space, the Hamiltonian $\hat{\mathcal{H}} = \sum_{\bm{k}} \hat{\psi}^\dagger_{\bm{k}} \hat{H}(\bm{k}) \hat{\psi}_{\bm{k}}$ is described by the $2 \times 2$ matrix\cite{Tanaka2025JPCM}
\begin{equation}
\hat{H}(\bm{k}) = \alpha_{\bm{k}}\sigma_0 - M_{\bm{k}}^p\sigma_z + \lambda (k_y \sigma_x - k_x \sigma_y) , \label{eq:1}
\end{equation}
where $\sigma_0$ is the $2 \times 2$ identity matrix, and $\bm{\sigma} = (\sigma_x, \sigma_y, \sigma_z)$ is the vector of Pauli matrices acting on the electron spin subspace. The spin-independent kinetic energy is defined as $\alpha_{\bm{k}} = \frac{\hbar^2 k^2}{2m}$, with $m$ representing the effective electron mass and $\bm{k} = (k_x, k_y)$ the 2D crystal momentum. The parameter $\lambda$ denotes the strength of the RSOC \cite{Bychkov1984}, which couples the electron's spin to its momentum and intrinsically breaks spatial inversion symmetry. Throughout this paper, $2\times2$ matrices in spin space carry a hat (e.g. $\hat{H}$, $\hat{G}$), with the exception of the Pauli matrices $\sigma_{0,x,y,z}$. Vectors are set in bold italic ($\bm{k}$, $\bm{R}$, $\bm{S}_i$, $\bm{G}$, $\bm{\mathcal{J}}_{\text{DM}}$), and scalars, including the tensor components $\mathcal{J}_{\ell m}$, in italic.

The unconventional magnetic order is entirely captured by the momentum-dependent exchange field $M_{\bm{k}}^p$. Unlike conventional ferromagnetism (where the exchange field is a constant) or $d$-wave altermagnetism (where the exchange field has an even-parity symmetry such as $k_x^2 - k_y^2$), the $p$-wave magnetic order is defined by an \textit{odd-parity} harmonic \cite{Tanaka2025JPCM, Maeda2024JPSJ} 
\begin{equation}
    M_{\bm{k}}^p = \frac{J}{k_F} (k_x \cos\beta + k_y \sin\beta), \label{eq:3}
\end{equation}
where $J$ defines the energy scale of the unconventional magnetic exchange, $k_F = \sqrt{2m\mathcal{E}_F/\hbar^2}$ is the unperturbed Fermi wavevector, $\mathcal{E}_F$ is the Fermi energy, and $\beta$ represents the angle between the $x$-axis and the polarization axis of the $p$-wave order. Setting $\beta = 0$ corresponds to a $p_x$-wave magnet, while $\beta = \pi/2$ corresponds to a $p_y$-wave magnet. It is crucial to note that $M_{-\bm{k}}^p = -M_{\bm{k}}^p$. This odd-parity inversion symmetry is the defining feature of $p$-wave magnets and, as we will show, fundamentally dictates the resulting spatial symmetries of the RKKY interaction tensor. While we focus our step-by-step main-text derivations on the $p$-wave case for clarity, the physical conclusions apply universally to higher-order odd-parity symmetries. In Appendix \ref{app:fwave}, we explicitly generalize this model to $f$-wave magnets, defined by an exchange field $M_{\bm{k}}^f \propto \cos 3(\theta - \beta)$.
We emphasize a fundamental distinction between this unconventional $p$-wave exchange field and the RSOC in Eq. (\ref{eq:1}). The $p$-wave term $M_{\bm{k}}^p \sigma_z$ is mathematically invariant under the standard time-reversal operator $\mathcal{T}$ in momentum space, bearing a structural resemblance to RSOC. However, its physical origin is fundamentally non-relativistic. In real space, $p$-wave magnets feature compensated non-collinear magnetic moments that break time-reversal symmetry $\mathcal{T}$ (which is not, by itself, a symmetry of the magnetic crystal) while carrying no net magnetization. The apparent time-reversal invariance of the low-energy Bloch bands is an emergent property protected by the composite symmetry of the lattice, typically $\mathcal{T}$ coupled with a half-unit-cell translation, $\tau$. Consequently, while the momentum-dependent splitting mathematically mimics relativistic spin-orbit coupling, it is driven by the Coulomb exchange interaction $J$. This emergent time-reversal invariance is not a shortcoming of the description but the very origin of the physics reported below. As shown in Sec.~\ref{subsec:momentumGF}, it is precisely what forces the spin-dependent part of the itinerant Green's function to be odd under $\bm{k}\to-\bm{k}$, and hence under $\bm{R}\to-\bm{R}$, which in turn dictates the parity-driven structure of the RKKY tensor. 

It is important to clarify the hierarchy of this continuum model. Equation (1) represents the pure phenomenological description of $p$-wave magnets, isolating the essential odd-parity momentum shift. However, derivations from specific microscopic non-collinear lattices or spin-spiral hosts typically yield an additional momentum-independent spin-flip exchange term. For example, for $\lambda=0$, Eq. (1) coincides with the sector-resolved low-energy expansion of the minimal lattice model of Ref. [\cite{Brekke2024}], $H_\eta(\boldsymbol{k})=(k^2/2m-\mu)\sigma_0+(\boldsymbol{\alpha}'\cdot\boldsymbol{k})\sigma_z+\eta J_{\rm sd}\sigma_x$ with $\eta=\pm$ labeling the two sectors, upon identifying $\boldsymbol{\alpha}'=-(J/k_F)(\cos\beta,\sin\beta)$, absorbing $\mu$ into $\mathcal{E}_F$, and omitting the lattice-specific term $J_{\rm sd}\to0$. This momentum-independent term $\eta J_{\rm sd}\sigma_x$ accounts for the spin-flip processes of the underlying inhomogeneous texture and renders the spin polarization of the bands partial. We omit it from our primary analytical model to keep the parity-driven RKKY mechanisms transparent. Being even in $\boldsymbol{k}$, it does not alter the odd parity of the $\sigma_z$ channel on which our conclusions rest, but, like the RSOC term retained here, it gaps the nodal lines. Its quantitative effect is rigorously addressed in Sec. \ref{subsec:materials} The continuum model thus captures the defining characteristics of $p$-wave magnetic order, namely an odd-parity, spin-conserving and non-relativistic momentum shift of the Fermi surfaces along an axis set by the magnetic texture, symmetry-enforced nodal lines, vanishing net magnetization and $\mathcal{T}\tau$ protection, while lattice-specific features (band warping, multiple bands, sublattice mixing) are addressed in Sec.\ref{subsec:materials}.

Diagonalizing the Hamiltonian in Eq.~(\ref{eq:1}) yields two spin-split energy branches 
\begin{equation}
    E_{\pm}(\bm{k}) = \alpha_{\bm{k}} \pm \sqrt{\lambda^2 k^2 + (M_{\bm{k}}^p)^2}, \label{eq:eigenvalues}
\end{equation}
where $k^2 = k_x^2 + k_y^2$. In the absence of RSOC ($\lambda = 0$), the energy bands are purely split by the $p$-wave exchange field. This results in a uniform rigid shift of the spin-up and spin-down Fermi surfaces in opposite directions along the momentum axis defined by $\beta$, preserving the circular shape of the Fermi contours and leaving the effective mass unrenormalized [Figs.~\ref{fig:rashba_interaction}(b) and \ref{fig:rashba_interaction}(c)]. However, the introduction of RSOC lifts the nodal degeneracies where $M_{\bm{k}}^p = 0$, leading to band hybridization and a rich, momentum-dependent spin texture [Figs.~\ref{fig:rashba_interaction}(d) and \ref{fig:rashba_interaction}(e)].

\begin{figure*}[t]
    \centering
    \includegraphics[width=\textwidth]{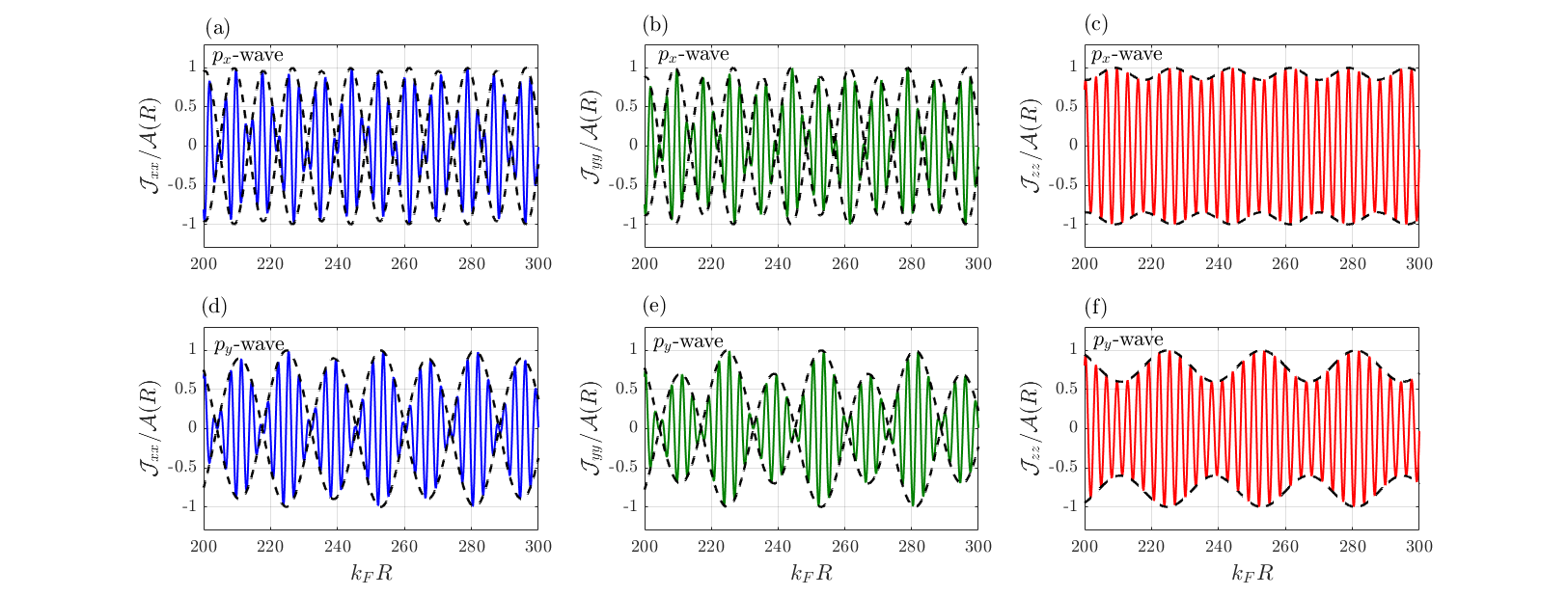}
    \caption{Spatial dependence of the normalized diagonal RKKY interaction components in a two-dimensional $p$-wave magnet subjected to RSOC. The top row [(a) to (c)] corresponds to a $p_x$-wave magnet ($\beta = 0$), while the bottom row [(d) to (f)] corresponds to a $p_y$-wave magnet ($\beta = \pi/2$). The columns display the in-plane Heisenberg components $\mathcal{J}_{xx}$ [(a), (d)] and $\mathcal{J}_{yy}$ [(b), (e)], and the out-of-plane Ising component $\mathcal{J}_{zz}$ [(c), (f)], all normalized by the isotropic spatial decay amplitude $\mathcal{A}(R)$. Black dashed lines trace the analytical spatial beating envelopes governed by the regularized momentum splitting $\kappa$. Retaining the full Rashba coupling breaks the in-plane spatial isotropy ($\mathcal{J}_{xx} \neq \mathcal{J}_{yy}$), inducing orientation-dependent constant shifts and distinct beating amplitudes parameterized by the impurity angle $\phi$. In contrast, the out-of-plane Ising interaction $\mathcal{J}_{zz}$ remains fundamentally decoupled from the macroscopic $p$-wave beating, oscillating predominantly at the shifted Fermi-surface radial wavevector $2k_F$, with only a minor RSOC-induced beating envelope of order $(k_R/\kappa)^2$. The plotted envelopes correspond to the dominant asymptotic expansion, neglecting sub-leading phase shifts of order $\mathcal{O}(k_R/k_F)$. The dimensionless parameters used are the effective $p$-wave exchange strength $Q/k_F = 0.20$, the effective RSOC strength $k_R/k_F = 0.05$, and an impurity separation angle $\phi = \pi/6$.}
    \label{fig:rkky_grid}
\end{figure*}

\subsection{Momentum-Space Retarded Green's Functions}
\label{subsec:momentumGF}
To evaluate the RKKY interaction mediated by the itinerant electrons in the $p$-wave magnet, we first compute the momentum-space retarded Green's function, defined as $\hat{G}(\bm{k}, \mathcal{E}) = [(\mathcal{E} + i\eta)\sigma_0 - \hat{H}(\bm{k})]^{-1}$, where $\mathcal{E}$ is the energy and $\eta$ is an infinitesimally small positive parameter ensuring causality\cite{bruus2004many}.

By inverting the $2 \times 2$ matrix $[(\mathcal{E} + i\eta)\sigma_0 - \hat{H}(\bm{k})]$, we retain the full relativistic spin-orbit coupling effects in the denominator. This treatment intrinsically captures the physical lifting of the degeneracy and the hybridized avoided-crossing bands near the $p$-wave nodal directions. Under this framework, the Green's function can be expressed compactly as $\hat{G}(\bm{k}, \mathcal{E}) = G_0(\bm{k}, \mathcal{E}) \sigma_0 + \bm{G}(\bm{k}, \mathcal{E}) \cdot \bm{\sigma}$, with the components 
\begin{align}
    G_0(\bm{k}, \mathcal{E}) &= \frac{\mathcal{E} + i\eta - \alpha_{\bm{k}}}{[\mathcal{E} + i\eta - \alpha_{\bm{k}}]^2 - \left[(M_{\bm{k}}^p)^2 + \lambda^2 k^2\right]}, \label{eq:11a_new} \\
    G_x(\bm{k}, \mathcal{E}) &= \frac{\lambda k_y}{[\mathcal{E} + i\eta - \alpha_{\bm{k}}]^2 - \left[(M_{\bm{k}}^p)^2 + \lambda^2 k^2\right]}, \label{eq:11b_new} \\
    G_y(\bm{k}, \mathcal{E}) &= \frac{-\lambda k_x}{[\mathcal{E} + i\eta - \alpha_{\bm{k}}]^2 - \left[(M_{\bm{k}}^p)^2 + \lambda^2 k^2\right]}, \label{eq:11c_new} \\
    G_z(\bm{k}, \mathcal{E}) &= \frac{-M_{\bm{k}}^p}{[\mathcal{E} + i\eta - \alpha_{\bm{k}}]^2 - \left[(M_{\bm{k}}^p)^2 + \lambda^2 k^2\right]}. \label{eq:11d_new}
\end{align}

Notice that because the kinetic energy $\alpha_{\bm{k}}$ and the Rashba term $\lambda^2 k^2$ are even functions of momentum ($\alpha_{-\bm{k}} = \alpha_{\bm{k}}$ and $k^2 = (-\bm{k})^2$), while $M_{\bm{k}}^p$, $k_x$, and $k_y$ are odd functions of momentum, the denominator is an even function of $\bm{k}$. Consequently, the scalar component $G_0(\bm{k}, \mathcal{E})$ is an \textit{even} function under spatial inversion. Conversely, the vector components $G_x, G_y$, and $G_z$ are all \textit{odd} functions of $\bm{k}$. This is a stark departure from even-parity $d$-wave altermagnets, where $G_z$ maintains even parity. We stress that these parity relations are not tied to the specific two-band form of Eq.~(\ref{eq:1}). They follow solely from the effective time-reversal invariance $\hat{H}(-\bm{k})=\hat{U}\hat{H}^{T}(\bm{k})\hat{U}^{-1}$ of Sec.~II.A, with $\hat{U}=\sigma_y$ here and $\hat{U}=\hat{U}_\tau\otimes\sigma_y$ for a multiband host, where $\hat{U}_\tau$ represents the half translation on the internal (sublattice or sector) indices. It implies $\hat{G}(-\bm{k},\mathcal{E})=\hat{U}\hat{G}^{T}(\bm{k},\mathcal{E})\hat{U}^{-1}$, which fixes the even (odd) behaviour of the scalar (spin) part of the propagator entering the RKKY trace, irrespective of the number of bands or of the form of the dispersion. In a multiband host the relation pairs the propagators of internal indices related by $\tau$ (Sec.~\ref{subsec:materials}). Only the explicit functional form of the components depends on the model.

\subsection{Real-Space Green's Functions}
\label{sec:RealSpaceGreenFunction}

The real-space Green's function governs the spatial propagation of electrons between the two magnetic impurities at $\bm{R}_1$ and $\bm{R}_2$. It is obtained via the two-dimensional Fourier transform, $\hat{G}(\bm{R}, \mathcal{E}) = \int \frac{d^2k}{(2\pi)^2} e^{i\bm{k}\cdot\bm{R}} \hat{G}(\bm{k}, \mathcal{E})$, where $\bm{R} = \bm{R}_2 - \bm{R}_1 = (R\cos\phi, R\sin\phi)$ is the separation vector. Given the parity relations established above, Fourier transforming to real space dictates that $G_0(-\bm{R}, \mathcal{E}) = G_0(\bm{R}, \mathcal{E})$ and $G_{x,y,z}(-\bm{R}, \mathcal{E}) = -G_{x,y,z}(\bm{R}, \mathcal{E})$.

To evaluate the integrals analytically, we utilize the polar coordinate representation $M_{\bm{k}}^p = \Delta_p k \cos(\theta - \beta)$, where $\Delta_p = J/k_F$. Because we retained the full relativistic Rashba term $\lambda^2 k^2$ in the denominator, the poles of the Green's functions seamlessly capture the hybridized, avoided-crossing bands. Utilizing the stationary-phase approximation (valid for large impurity separations $k_F R \gg 1$), the angular integration is dominated by the axis of impurity separation $\theta = \phi$. The resulting oscillatory behavior is dictated by the hybridized wavevectors $k_\pm = k_0 \pm \kappa$, where
\begin{equation}
    k_0 = \sqrt{\frac{2m\mathcal{E}}{\hbar^2} + \kappa^2}, \quad \kappa = \sqrt{k_M^2 + k_R^2}. \label{eq:8_new}
\end{equation}
Here, $k_M = Q \cos(\phi - \beta)$ with $Q = \frac{m\Delta_p}{\hbar^2}$ acts as the anisotropic magnetic modulation wavevector, and $k_R = m\lambda/\hbar^2$ is the intrinsic Rashba wavevector. The term $\kappa$ represents the regularized spin splitting. We note that all host-induced length scales, namely the projected shift $k_M$, the regularized splitting $\kappa$, the beating wavelength $\pi/\kappa$, and the nodal crossover of Sec.~\ref{subsec:decay}, depend on the impurity orientation $\phi$ and on the polarization axis $\beta$ only through the relative angle $\phi-\beta$. The isolated factors of $\sin\phi$ and $\cos\phi$ that appear below in the Rashba-induced components are purely geometric. They originate from the projection of the spin-momentum locking $\lambda(k_y\sigma_x-k_x\sigma_y)$ onto the stationary direction $\bm{k}\parallel\bm{R}$ and express the bond-dependent interaction in the fixed laboratory spin frame. We therefore keep $\phi$ and $\beta$ explicit, as they are controlled independently in experiment ($\phi$ by the placement of the adatoms, $\beta$ by the orientation of the magnetic order).

Applying the Cauchy residue theorem to perform the momentum integration, we extract the analytical expressions in terms of Hankel functions of the first kind (The details are given in Appendix \ref{Appendix A}). The diagonal components yield 
\begin{align}
    G_0(\bm{R}, \mathcal{E}) &= -\frac{im}{4\hbar^2} \left[ H_0^{(1)}(k_+R) + H_0^{(1)}(k_-R) \right], \label{eq:9_new} \\
    G_z(\bm{R}, \mathcal{E}) &= -\frac{im}{4\hbar^2} \left( \frac{k_M}{\kappa} \right) \left[ H_0^{(1)}(k_+R) - H_0^{(1)}(k_-R) \right]. \label{eq:10_new}
\end{align}
Notice that $G_z$ acquires the dimensionless spin-projection factor $k_M/\kappa$ directly from the matrix inversion, forcing the out-of-plane magnetic propagation to vanish continuously at the $p$-wave nodal lines where $k_M \to 0$.

For the off-diagonal Rashba components, the spatial derivatives $-i\partial_y$ and $i\partial_x$ pull down momentum factors and raise the order of the Hankel functions from $H_0^{(1)}$ to $H_1^{(1)}$. The $p$-wave symmetry enforces a relative sign difference between the two spin branches 
\begin{align}
    G_x(\bm{R}, \mathcal{E}) &= \frac{ m \sin\phi}{4\hbar^2}\frac{k_R}{\kappa}\frac{1}{k_0} \left[ k_+ H_1^{(1)}(k_+R) - k_- H_1^{(1)}(k_-R) \right], \label{eq:11_new} \\
    G_y(\bm{R}, \mathcal{E}) &=\frac{- m \cos\phi}{4\hbar^2}\frac{k_R}{\kappa}\frac{1}{k_0} \left[ k_+ H_1^{(1)}(k_+R) - k_- H_1^{(1)}(k_-R) \right]. \label{eq:12_new}
\end{align}
\noindent These expressions inherently eliminate nodal divergences, providing a completely controlled analytical framework for the indirect exchange everywhere, including on the $p$-wave nodal lines.

\begin{figure}
    \centering
    \includegraphics[width=0.48\textwidth]{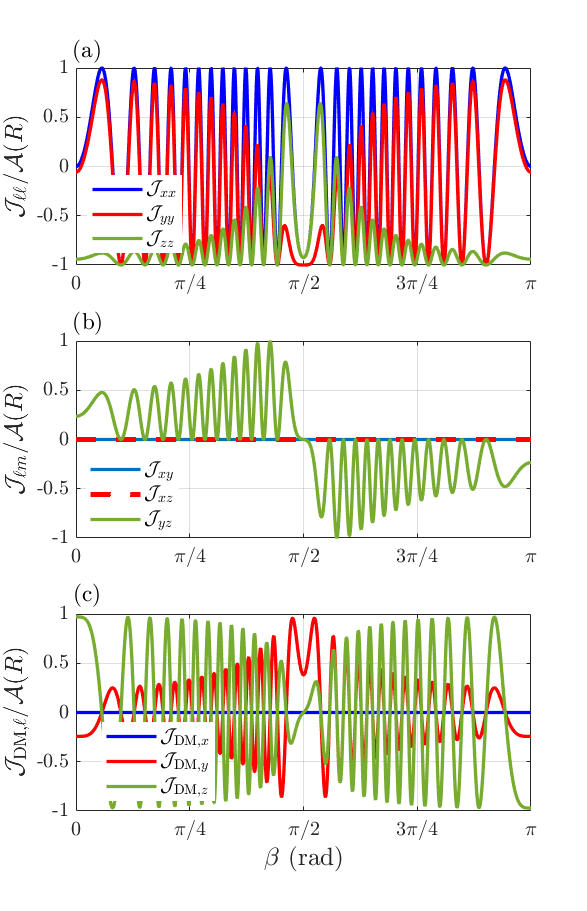}
    \caption{Evolution of the normalized RKKY interaction tensor components as a function of the $p$-wave polarization angle $\beta$, evaluated for two magnetic impurities aligned along the $x$-axis ($\phi = 0$). The panels display (a) the diagonal Heisenberg and Ising interactions, (b) the symmetric off-diagonal interactions, and (c) the antisymmetric DM components. Aligning the impurity bond precisely with the $x$-axis imposes a geometric symmetry ($\sin\phi = 0$) that identically quenches the $\mathcal{J}_{xy}$, $\mathcal{J}_{xz}$, and $\mathcal{J}_{\text{DM},x}$ components across all polarization angles. In panel (b) the vanishing $\mathcal{J}_{xy}$ and $\mathcal{J}_{xz}$ are drawn as solid and dashed lines, respectively, so that both remain visible. As the $p$-wave axis rotates, the projected momentum shift $k_M = Q\cos(\beta)$ highly tunes the effective beating wavevector, driving pronounced angular oscillations. At the orthogonal nodal configuration ($\beta = \pi/2$), the projected shift vanishes ($k_M = 0$). Consequently, the macroscopic nonrelativistic out-of-plane chiral component $\mathcal{J}_{\text{DM},z}$ vanishes, while the relativistic Rashba-induced components ($\mathcal{J}_{\text{DM},y}$ and $\mathcal{J}_{yz}$) exhibit finite, regularized crossover peaks bounded by the intrinsic Rashba gap $\kappa = k_R$. Panel (a) visually reconfirms the parity-driven decoupling. The out-of-plane Ising interaction $\mathcal{J}_{zz}$ is structurally insulated from the massive $p$-wave beating that heavily modulates the in-plane Heisenberg components. The fixed dimensionless parameters are $k_F R = 255.25$, $Q/k_F = 0.20$, and $k_R/k_F = 0.05$.}
    \label{fig:RKKY_beta}
\end{figure}

\subsection{RKKY Exchange Tensor}
The localized impurity spins interact with the spin density of the itinerant electrons via the $s$-$d$ exchange Hamiltonian, $\mathcal{H}_{\text{int}} = -J_{\text{imp}} \sum_{i=1,2} \bm{S}_i \cdot \bm{\sigma}(\bm{R}_i)$, where $J_{\text{imp}}$ denotes the local exchange coupling strength. Applying standard second-order perturbation theory at zero temperature, the effective RKKY interaction is derived from the trace over the itinerant electron spin degree of freedom \cite{Zhu2011, Chesi2010PRB,Yarmohammadi2025,Amundsen2024,Yarmohammadi2026} 
\begin{equation}
    \mathcal{H}_{\text{RKKY}} = \frac{-J_{\text{imp}}^2}{\pi} \text{Im} \int_{-\infty}^{\mathcal{E}_F} d\mathcal{E} \, \text{Tr} \Big[ (\bm{S}_1 \cdot \bm{\sigma}) \hat{G}(\bm{R}) (\bm{S}_2 \cdot \bm{\sigma}) \hat{G}(-\bm{R}) \Big].
    \label{eq:H_RKKY}
\end{equation}

Because the scalar Green's function is even under spatial inversion, $G_0(-\bm{R}) = G_0(\bm{R})$, while the vector components are odd, $G_{x,y,z}(-\bm{R}) = -G_{x,y,z}(\bm{R})$, the total interaction systematically decouples into a symmetric tensor $\mathcal{J}_{\ell m}$ and an antisymmetric DM vector $\bm{\mathcal{J}}_{\text{DM}} $\cite{Moriya1960}
\begin{equation}
\begin{array}{rl}
    \mathcal{H}_{\text{RKKY}} & = \sum_{\ell=x,y,z} \mathcal{J}_{\ell\ell}(\bm{R}) S_1^\ell S_2^\ell
    \\
    &\\
    & + \sum_{\ell<m} \mathcal{J}_{\ell m}(\bm{R}) \left[ S_1^\ell S_2^m + S_1^m S_2^\ell \right]\\
    & \\& 
    + \bm{\mathcal{J}}_{\text{DM}}(\bm{R}) \cdot (\bm{S}_1 \times \bm{S}_2).
\end{array}
\end{equation}

By expanding the trace over the Pauli matrices, the diagonal components, which represent the conventional Heisenberg and Ising exchange interactions, evaluate to
\begin{align}
    \mathcal{J}_{xx}(\bm{R}) &= -\frac{2J_{\text{imp}}^2}{\pi} \text{Im} \int_{-\infty}^{\mathcal{E}_F} d\mathcal{E} \Big[ G_0^2 - G_x^2 + G_y^2 + G_z^2 \Big], \label{eq:Jxx} \\
    \mathcal{J}_{yy}(\bm{R}) &= -\frac{2J_{\text{imp}}^2}{\pi} \text{Im} \int_{-\infty}^{\mathcal{E}_F} d\mathcal{E} \Big[ G_0^2 + G_x^2 - G_y^2 + G_z^2 \Big], \label{eq:Jyy} \\
    \mathcal{J}_{zz}(\bm{R}) &= -\frac{2J_{\text{imp}}^2}{\pi} \text{Im} \int_{-\infty}^{\mathcal{E}_F} d\mathcal{E} \Big[ G_0^2 + G_x^2 + G_y^2 - G_z^2 \Big]. \label{eq:Jzz}
\end{align}
These equations reveal a profound physical signature of the $p$-wave state. In conventional ferromagnets or even-parity $d$-wave altermagnets, $G_z$ is an even function of space, meaning $G_z(\bm{R})G_z(-\bm{R}) = +G_z^2(\bm{R})$. However, because the $p$-wave exchange field is an odd function of momentum, $G_z$ acquires an odd parity in real space, yielding $G_z(\bm{R})G_z(-\bm{R}) = -G_z^2(\bm{R})$. This parity-driven sign reversal fundamentally alters the interference between the charge ($G_0^2$) and magnetic ($G_z^2$) propagation channels. It forces a strict $-G_z^2$ contribution in the out-of-plane Ising interaction ($\mathcal{J}_{zz}$) and a $+G_z^2$ contribution in the in-plane components. We note that the in-plane equality $\mathcal{J}_{xx}=\mathcal{J}_{yy}$ holds only at zeroth order in $\lambda$. The combinations $-G_x^2+G_y^2$ and $+G_x^2-G_y^2$ in Eqs.~(\ref{eq:Jxx}) and~(\ref{eq:Jyy}) coincide at $\mathcal{O}(\lambda^0)$ but differ at $\mathcal{O}(\lambda^2)$, meaning a finite RSOC weakly breaks the in-plane isotropy at the next order.

Our trace keeps the symmetric off-diagonal interactions, which map to the cross-products of the vector Green's functions
\begin{align}
    \mathcal{J}_{xy}(\bm{R}) &= \frac{4 J_{\text{imp}}^2}{\pi} \text{Im} \int_{-\infty}^{\mathcal{E}_F} d\mathcal{E} \Big[ G_x G_y \Big], \label{eq:Jxy} \\
    \mathcal{J}_{yz}(\bm{R}) &= \frac{4 J_{\text{imp}}^2}{\pi} \text{Im} \int_{-\infty}^{\mathcal{E}_F} d\mathcal{E} \Big[ G_y G_z \Big], \label{eq:Jyz} \\
    \mathcal{J}_{xz}(\bm{R}) &= \frac{4 J_{\text{imp}}^2}{\pi} \text{Im} \int_{-\infty}^{\mathcal{E}_F} d\mathcal{E} \Big[ G_x G_z \Big]. \label{eq:Jxz}
\end{align}
The $G_x$ and $G_y$ are proportional to $\lambda$, so $\mathcal{J}_{xy}$ is $\mathcal{O}(\lambda^2)$. By contrast, $\mathcal{J}_{yz}$ and $\mathcal{J}_{xz}$ are products of one Rashba propagator and the $p$-wave $G_z$ propagator. Therefore, they scale as $\mathcal{O}(\lambda)$, the same formal order as the in-plane DM components. 

Finally, the RSOC intrinsically breaks spatial inversion symmetry and couples with the scalar propagation $G_0$ to generate the antisymmetric DM vector with components of 
\begin{align}
    \mathcal{J}_{\text{DM},x}(\bm{R}) &= \frac{4 J_{\text{imp}}^2}{\pi} \text{Im} \int_{-\infty}^{\mathcal{E}_F} d\mathcal{E} \Big[ i G_0 G_x \Big], \label{eq:JDMx} \\
    \mathcal{J}_{\text{DM},y}(\bm{R}) &= \frac{4 J_{\text{imp}}^2}{\pi} \text{Im} \int_{-\infty}^{\mathcal{E}_F} d\mathcal{E} \Big[ i G_0 G_y \Big], \label{eq:JDMy} \\
    \mathcal{J}_{\text{DM},z}(\bm{R}) &= \frac{4 J_{\text{imp}}^2}{\pi} \text{Im} \int_{-\infty}^{\mathcal{E}_F} d\mathcal{E} \Big[ i G_0 G_z \Big]. \label{eq:JDMz}
\end{align}
These DM components favor a non-collinear, chiral twisting of the impurity spins. The in-plane components $\mathcal{J}_{\text{DM},x/y}$ are directly proportional to the Rashba coupling $\lambda$. In stark contrast, the out-of-plane component $\mathcal{J}_{\text{DM},z} \propto iG_0 G_z$ is built exclusively from the scalar and $p$-wave channels alone, meaning it is present even at $\lambda = 0$. The factor $iG_0 G_z$ inherits the complex Hankel envelope $H_0^{(1)}(k_0R)$ shared by $G_0$ and $G_z$, ensuring a finite imaginary part. We note that the overall sign of $\bm{\mathcal{J}}_{\text{DM}}$ reflects the impurity-labeling convention (namely the orientation of $\bm{R}=\bm{R}_2-\bm{R}_1$) and is immaterial to the magnitudes and spatial profiles reported here (For more details, have a look on Appendix \ref{Appendix B}). 

\begin{figure*}[t]
    \centering
    \includegraphics[width=\textwidth]{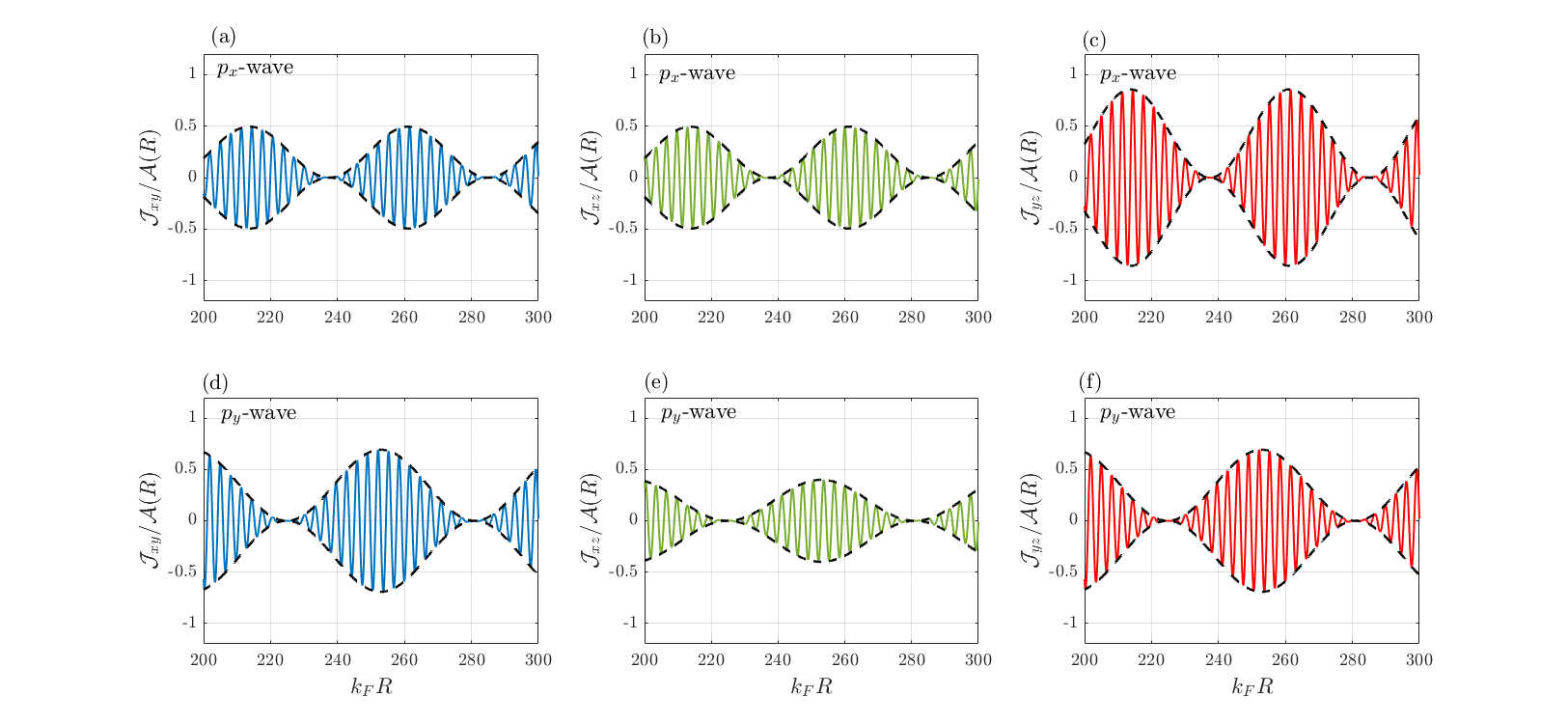}
    \caption{Spatial dependence of the normalized symmetric off-diagonal RKKY interaction components in a two-dimensional $p$-wave magnet subjected to RSOC. The top row [(a) to (c)] corresponds to a $p_x$-wave magnet ($\beta = 0$), while the bottom row [(d) to (f)] corresponds to a $p_y$-wave magnet ($\beta = \pi/2$). The columns display the in-plane component $\mathcal{J}_{xy}$ [(a), (d)], and the out-of-plane cross components $\mathcal{J}_{xz}$ [(b), (e)] and $\mathcal{J}_{yz}$ [(c), (f)], all normalized by the isotropic spatial decay amplitude $\mathcal{A}(R)$. Black dashed lines trace the analytical spatial beating envelopes governed by the regularized momentum splitting $\kappa$. Unlike the diagonal Heisenberg channels, the symmetric off-diagonal terms exhibit a pure $(1 - \cos(2\kappa R))$ beating envelope with no constant offset, resulting in perfect nodal zero-crossings where the anisotropic exchange is completely quenched. The rotation of the $p$-wave order parameter from $p_x$ to $p_y$ highly tunes both the beating wavelength and the maximum coupling amplitudes by modifying the projected momentum shift $k_M$. The dimensionless parameters used are the effective $p$-wave exchange strength $Q/k_F = 0.05$, the effective RSOC strength $k_R/k_F = 0.05$, and an impurity separation angle $\phi = \pi/6$. Sub-leading relativistic phase shifts of order $\mathcal{O}(\kappa/k_F)$ are neglected to isolate the dominant macroscopic beating.}
    \label{fig:OffDiagonal}
\end{figure*}
\begin{figure*}[t]
    \centering
    \includegraphics[width=\textwidth]{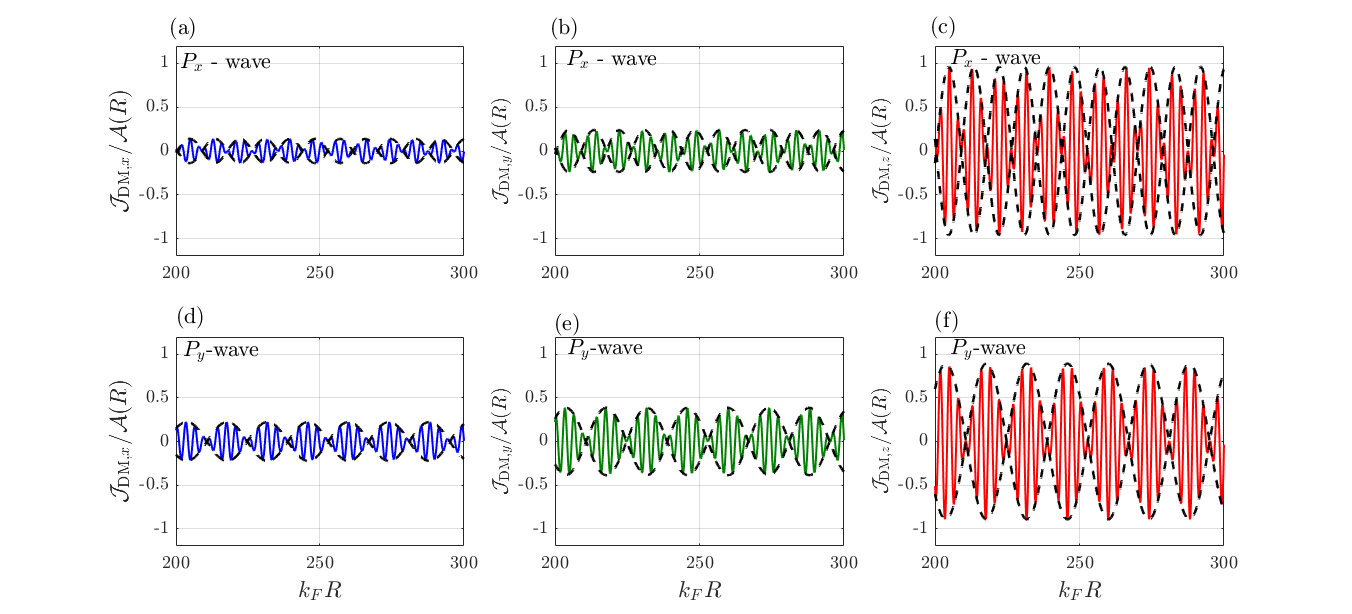}
    \caption{Spatial dependence of the normalized DM interaction components in a two-dimensional $p$-wave magnet subjected to RSOC. The top row [(a) to (c)] corresponds to a $p_x$-wave magnet ($\beta = 0$), while the bottom row [(d) to (f)] corresponds to a $p_y$-wave magnet ($\beta = \pi/2$). The columns display the in-plane chiral components $\mathcal{J}_{\text{DM},x}$ [(a), (d)] and $\mathcal{J}_{\text{DM},y}$ [(b), (e)], and the out-of-plane component $\mathcal{J}_{\text{DM},z}$ [(c), (f)], all normalized by the isotropic spatial decay amplitude $\mathcal{A}(R)$. Black dashed lines trace the analytical spatial beating envelopes governed by the regularized momentum splitting $\kappa$. The antisymmetric DM vector exhibits a pure, zero-centered $\sin(2\kappa R)$ spatial beating. Rotating the $p$-wave order parameter from $p_x$ to $p_y$ modifies the projected momentum shift $k_M$, tunably reshaping the spatial beating wavelengths and redistributing the relative magnitudes of the in-plane components. The dimensionless parameters used are $Q/k_F = 0.20$, $k_R/k_F = 0.05$, and an impurity separation angle $\phi = \pi/6$. Sub-leading relativistic corrections of order $\mathcal{O}(k_R/k_F)$ are neglected to isolate the dominant macroscopic beating.}
    \label{fig:DMcomponents}
\end{figure*}

\section{RESULTS AND DISCUSSION}
\label{sec:results}

The shifted-contour forms of $G_0$ and $G_z$ reveal a striking physical decoupling in the RKKY responses. The results below are long-distance asymptotic expressions, valid for $k_F R \gg 1$. To obtain the full long-range asymptotic limit, we apply the large-argument expansion of the Hankel functions, $H_\nu^{(1)}(z) \approx \sqrt{\frac{2}{\pi z}} e^{i(z - \nu\pi/2 - \pi/4)}$. The energy integration over the Fermi sea is therefore rigorously dominated by the poles evaluated at the Fermi level, $k_\pm = k_0(\mathcal{E}_F) \pm \kappa$, continuously capturing the avoided-crossing bands across all spatial orientations. Dropping subleading asymptotic terms, we calculate the interaction tensor components to illustrate the unique spatial signatures and the robust non-collinear chiral twisting inherent to the $p$-wave state.

\subsection{Parity-Driven Role Reversal in the Ising and Heisenberg Interactions}
\label{subsec:roles}

By evaluating the Green's functions across the energy bands, we extract the diagonal elements of the exchange tensor. To isolate the dominant spatial beating envelopes, we apply the realistic material limit $k_R \ll k_F$. This safely neglects sub-leading amplitude corrections of $\mathcal{O}(k_R^2/k_F^2)$ and relativistic spatial phase shifts of $\mathcal{O}(k_R/k_F)$ (we direct the reader to Appendix \ref{Appendix C} for the fully unapproximated expressions). Integrating over the Fermi sea yields the regularized, analytically closed-form Heisenberg and Ising RKKY interactions
\begin{widetext}
  \begin{align}
\mathcal{J}_{xx}(\bm{R}) \simeq & -\mathcal{A}(R) \left[ \frac{k_R^2 \sin^2\phi}{\kappa^2} + \frac{k_M^2 + k_R^2 \cos^2\phi}{\kappa^2}\cos(2\kappa R) \right] \sin(2k_FR), \label{eq:J_xx_full} \\
\mathcal{J}_{yy}(\bm{R}) \simeq & -\mathcal{A}(R) \left[ \frac{k_R^2 \cos^2\phi}{\kappa^2} + \frac{k_M^2 + k_R^2 \sin^2\phi}{\kappa^2}\cos(2\kappa R) \right] \sin(2k_FR), \label{eq:J_yy_full} \\
\mathcal{J}_{zz}(\bm{R}) \simeq & -\mathcal{A}(R) \left[ \frac{k_M^2}{\kappa^2} + \frac{k_R^2}{\kappa^2}\cos(2\kappa R) \right] \sin(2k_F R), \label{eq:J_zz_full}
\end{align}  
\end{widetext}

where $\mathcal{A}(R) = \frac{J_{\text{imp}}^2 m}{2\pi^2\hbar^2 R^2}$ is the standard isotropic RKKY spatial decay amplitude, showing the expected $1/R^2$ scaling for a 2D electron gas \cite{BealMonod1987}. 

Equations (\ref{eq:J_xx_full}) to (\ref{eq:J_zz_full}) reveal that retaining the full Rashba coupling systematically breaks the in-plane spatial isotropy ($\mathcal{J}_{xx} \neq \mathcal{J}_{yy}$). As depicted in the first two columns of Fig.~\ref{fig:rkky_grid}, the Heisenberg components exhibit highly directional constant shifts and distinct beating amplitudes parameterized by the impurity orientation angle $\phi$. In stark contrast, the out-of-plane Ising component $\mathcal{J}_{zz}$ [Figs.~\ref{fig:rkky_grid}(c) and \ref{fig:rkky_grid}(f)] is structurally dominated by the non-beating constant shift $k_M^2/\kappa^2$, exhibiting only a minor Rashba-induced spatial beating envelope proportional to $k_R^2/\kappa^2$. 

To transparently uncover the theoretical origin of this magnetic decoupling, we analyze the system away from the nodal lines, where the $p$-wave exchange dominates the relativistic coupling ($k_M \gg k_R$). In this limit, $\kappa \to k_M$, the in-plane isotropy is restored, and the interaction tensor elegantly collapses to
\begin{align}
\mathcal{J}_{xx/yy}(\bm{R}) \simeq & -\mathcal{A}(R) \cos(2 k_M R) \sin(2k_FR),  \label{eq:J_xx_simp}\\
\mathcal{J}_{zz}(\bm{R}) \simeq & -\mathcal{A}(R) \sin(2k_F R ). \label{eq:J_zz_simp} 
\end{align}

These simplified limits highlight a fundamental parity-driven ``role reversal'' when compared to even-parity $d$-wave altermagnets. In a $d$-wave system, the momentum-dependent exchange field is an even function in momentum space ($M_{-\bm{k}} = M_{\bm{k}}$). Consequently, its spin propagator $G_z(\bm{R})$ is an even function in real space, meaning $G_z(\bm{R})G_z(-\bm{R}) = +G_z^2(\bm{R})$. In the RKKY trace, this positive sign forces the out-of-plane Ising interaction (proportional to $G_0^2 + G_z^2$) to couple strongly to the spatial beating, while the in-plane interactions (proportional to $G_0^2 - G_z^2$) cancel the beating, deviating only slightly from a normal metal \cite{Amundsen2024}. 

Here, in a $p$-wave magnet, the opposite parity structure dictates the physics. Because the $p$-wave exchange field is strictly odd in momentum space ($M_{-\bm{k}}^p = -M_{\bm{k}}^p$), the resulting real-space spin propagator inherits an odd parity, $G_z(-\bm{R}) = -G_z(\bm{R})$. When traversing the indirect exchange loop between two impurities, this parity flips the interference sign, yielding $G_z(\bm{R})G_z(-\bm{R}) = -G_z^2(\bm{R})$. This microscopic sign reversal fundamentally re-routes the spatial beating. The out-of-plane Ising term is now governed by $G_0^2 - G_z^2$, stripping it of the $k_M$ modulation, while the in-plane Heisenberg channels are governed by $G_0^2 + G_z^2$, completely absorbing the robust $\cos(2k_M R)$ envelope. 

As a direct consequence, the out-of-plane Ising interaction $\mathcal{J}_{zz}$ escapes the macroscopic beating entirely. As visibly demonstrated in Figs.~\ref{fig:rkky_grid}(c) and \ref{fig:rkky_grid}(f), it oscillates at the shifted Fermi-surface radial wave number denoted by $2k_F$, acting effectively ``blind'' to the underlying anisotropic $p$-wave magnetic order.

To continuously visualize this striking parity-driven decoupling as the magnetic order is rotated, Fig.~\ref{fig:RKKY_beta}(a) tracks the diagonal interactions as a function of the $p$-wave polarization angle $\beta$ for a fixed impurity pair aligned along the $x$-axis ($\phi=0$). As $\beta$ is tuned, the projected momentum shift $k_M = Q\cos(\beta)$ sweeps from its maximum to zero. Consequently, the in-plane Heisenberg components ($\mathcal{J}_{xx}, \mathcal{J}_{yy}$) exhibit rapid, pronounced macroscopic beating. In complete contrast, the out-of-plane Ising component $\mathcal{J}_{zz}$ remains fundamentally insulated from this modulation, maintaining a near-constant amplitude and acquiring only a weak relativistic ripple near the nodal configuration ($\beta = \pi/2$), providing visual confirmation of its structural decoupling from the $p$-wave order.

\subsection{Symmetric Off-Diagonal Exchange Interactions}

The RKKY tensor contains symmetric off-diagonal anisotropies ($\mathcal{J}_{xy}$, $\mathcal{J}_{xz}$, $\mathcal{J}_{yz}$) that fundamentally drive non-collinear spin twisting. By retaining the fully regularized momentum splitting $\kappa$ and neglecting sub-leading phase shifts of order $\mathcal{O}(\kappa/k_F)$, the closed-form leading stationary-phase terms evaluate to
\begin{align}
\mathcal{J}_{xy}(\bm{R}) &\simeq \mathcal{A}(R) \frac{k_R^2}{2\kappa^2} \sin(2\phi) \big[1 - \cos(2\kappa R)\big] \sin(2k_F R), \label{eq:J_xy_sym} \\
\mathcal{J}_{xz}(\bm{R}) &\simeq -\mathcal{A}(R) \frac{k_R k_M}{\kappa^2} \sin\phi \big[1 - \cos(2\kappa R)\big] \sin(2k_F R), \label{eq:J_xz_sym} \\
\mathcal{J}_{yz}(\bm{R}) &\simeq \mathcal{A}(R) \frac{k_R k_M}{\kappa^2} \cos\phi \big[1 - \cos(2\kappa R)\big] \sin(2k_F R), \label{eq:J_yz_sym}
\end{align}
These terms vanish only in special geometries or limits where the corresponding trigonometric prefactors vanish. Thus, the symmetric tensor is purely diagonal only at $\lambda=0$ (or for special high-symmetry impurity orientations). The full expressions are obtained in Appendix C. 

As depicted in Fig.~\ref{fig:OffDiagonal}, the spatial profiles of these symmetric off-diagonal components reveal a striking departure from the diagonal Heisenberg channels. While the in-plane Heisenberg components ($\mathcal{J}_{xx}, \mathcal{J}_{yy}$) contain a constant, unmodulated background offset, the symmetric off-diagonal interactions are enveloped by a pure $\big[1 - \cos(2\kappa R)\big]$ spatial beating. Consequently, their spatial envelopes reach the zero-axis, indicating that these anisotropic couplings are pure interference phenomena born from the hybridized spin-orbit-coupled bands. At precise, regular spatial distances where $2\kappa R = 2n\pi$, the symmetric off-diagonal exchange is completely quenched. 

Furthermore, these components illustrate the formal hierarchy of the relativistic Rashba coupling. The in-plane component $\mathcal{J}_{xy}$ is second-order in the RSOC ($\mathcal{O}(\lambda^2)$), as it requires the cross-product of two Rashba propagators ($G_x G_y$). By contrast, the out-of-plane cross-components $\mathcal{J}_{xz}$ and $\mathcal{J}_{yz}$ scale linearly with the RSOC ($\mathcal{O}(\lambda)$), emerging from the interference between one Rashba and one $p$-wave propagator. In a strict asymptotic limit where $\lambda \to 0$, $\mathcal{J}_{xy}$ would be neglected compared to $\mathcal{J}_{xz}$ and $\mathcal{J}_{yz}$. However, when the relativistic momentum shift is comparable to the macroscopic $p$-wave splitting ($k_R \sim k_M$), this strict weak-RSOC hierarchy breaks down. In this regime, all three off-diagonal components exhibit comparable macroscopic magnitudes, as clearly demonstrated in Figs.~\ref{fig:OffDiagonal}(a)-(c), meaning the full noncollinear response cannot be captured by the DM vector alone.

Finally, a comparison between the $p_x$-wave ($\beta=0$, top row) and $p_y$-wave ($\beta=\pi/2$, bottom row) configurations in Fig.~\ref{fig:OffDiagonal} highlights the tunability of these interactions. Rotating the $p$-wave polarization axis modifies the projected magnetic wavevector $k_M = Q\cos(\phi-\beta)$. Because both the overall interaction amplitudes (proportional to $1/\kappa^2$ or $k_M/\kappa^2$) and the spatial beating wavelength (dictated by $\kappa$) depend on $k_M$, changing the $p$-wave orientation relative to the impurity bond reshapes the spatial hierarchy of the off-diagonal exchange. As shown in Fig. \ref{fig:RKKY_beta} (b), this provides a highly controllable mechanism for engineering direction-dependent magnetic frustration and non-collinear spin textures in unconventional $p$-wave magnets.

\subsection{Dzyaloshinskii-Moriya Interactions and Nodal Chiral Enhancement}

The introduction of RSOC breaks spatial inversion symmetry and gives rise to the antisymmetric DM interaction \cite{FertLevy1980}. By evaluating the cross-products of the real-space Green's functions while retaining the fully hybridized energy denominator, we derive the analytically closed-form asymptotic DM tensor components in Appendix \ref{Appendix C}. Neglecting sub-leading relativistic phase shifts of order $\mathcal{O}(k_R/k_F)$, the expressions are
\begin{align}
\mathcal{J}_{\text{DM},x}(\bm{R}) &\simeq -\mathcal{A}(R)\,\frac{k_R}{\kappa}\,\sin\phi\,\sin(2\kappa R)\,\sin(2k_F R), \label{eq:J_DMx_reg} \\
\mathcal{J}_{\text{DM},y}(\bm{R}) &\simeq \mathcal{A}(R)\,\frac{k_R}{\kappa}\,\cos\phi\,\sin(2\kappa R)\,\sin(2k_F R), \label{eq:J_DMy_reg} \\
\mathcal{J}_{\text{DM},z}(\bm{R}) &\simeq -\mathcal{A}(R)\,\frac{k_M}{\kappa}\,\sin(2\kappa R)\,\sin(2k_F R), \label{eq:J_DMz_reg}
\end{align}

The spatial profiles of these chiral interactions are illustrated in Fig.~\ref{fig:DMcomponents}. In stark contrast to the diagonal Heisenberg components and the symmetric off-diagonal $\big[1-\cos(2\kappa R)\big]$ envelopes, the antisymmetric DM vector is governed by a pure, zero-centered $\sin(2\kappa R)$ spatial beating. At distances where $2\kappa R = n\pi$, the chiral coupling vanishes.

Furthermore, Fig.~\ref{fig:DMcomponents} visually underscores the profound physical hierarchy of the $p$-wave chiral exchange. The out-of-plane component $\mathcal{J}_{\text{DM},z}$ is driven primarily by the macroscopic $p$-wave momentum shift. Its physical origin is entirely nonrelativistic and is transparent in the $\lambda=0$ limit, where the $p$-wave exchange shifts the two spin sectors by opposite momenta, which is equivalent to a position-dependent spin rotation about $\hat{z}$. For a single impurity pair, this can be viewed as a twisted Heisenberg interaction written in the fixed laboratory spin frame, and it naturally produces robust, $\mathcal{O}(1)$ macroscopic interaction amplitudes. Conversely, the in-plane components $\mathcal{J}_{\text{DM},x}$ and $\mathcal{J}_{\text{DM},y}$ are relativistic Rashba features. They scale proportionally to $k_R/\kappa$ and are consequently much weaker in the physically relevant limit where $Q \gg k_R$. In extended impurity arrays, the combination of these bond-dependent twists and the pinning of the spin frame by RSOC makes the resulting spiral tendencies observable and distinguishes them from a conventional spin-orbit-induced DM interaction.

Comparing the $p_x$-wave ($\beta=0$) and $p_y$-wave ($\beta=\pi/2$) configurations in Fig.~\ref{fig:DMcomponents} highlights the high degree of directional tunability in these materials. Rotating the $p$-wave polarization angle $\beta$ drastically modifies the projected momentum shift $k_M = Q\cos(\phi-\beta)$. Because both the interaction amplitudes and the beating wavevector $\kappa$ depend intimately on $k_M$, changing the $p$-wave orientation tunably reshapes the spatial beating wavelength $\lambda_{\text{beat}} = \pi/\kappa$ and redistributes the relative magnitudes of the chiral components.

This mechanism is fundamentally distinct from the RKKY response in even-parity $d$-wave altermagnets. In $d$-wave systems, the even-parity momentum splitting primarily imposes highly anisotropic directional beating onto the out-of-plane Ising channel, while the chiral twisting remains predominantly relativistic. Here, the odd-parity $p$-wave symmetry induces a parity-driven "role reversal." The out-of-plane Ising interaction is decoupled from the macroscopic beating entirely, while a massive, nonrelativistic out-of-plane DM interaction emerges as the primary driver. This identifies $p$-wave magnets as uniquely suited platforms for engineering highly anisotropic, non-collinear spin spirals tunable by electric-field control of the Rashba coupling or rotation of the crystalline $p$-wave axis.

The full dynamical crossover of this chiral hierarchy is mapped in Fig.~\ref{fig:RKKY_beta}(c). As the $p$-wave polarization $\beta$ is rotated continuously toward the nodal configuration ($\beta=\pi/2$), the system undergoes a profound inversion of its dominant chiral axis. The macroscopic, nonrelativistic out-of-plane component $\mathcal{J}_{\text{DM},z}$ scales directly with $k_M$ and therefore smoothly decays to zero at the node. Concurrently, the relativistic in-plane component $\mathcal{J}_{\text{DM},y}$ achieves a finite, regularized maximum bounded by the intrinsic Rashba gap ($\kappa \to k_R$). This continuous angular crossover provides a highly controllable mechanism to dynamically sweep the dominant axis of chiral twisting from out-of-plane to purely in-plane, confirming the robust potential of $p$-wave architectures in orientation-driven spintronics.

\begin{figure}
    \centering
    \includegraphics[width=0.45\textwidth]{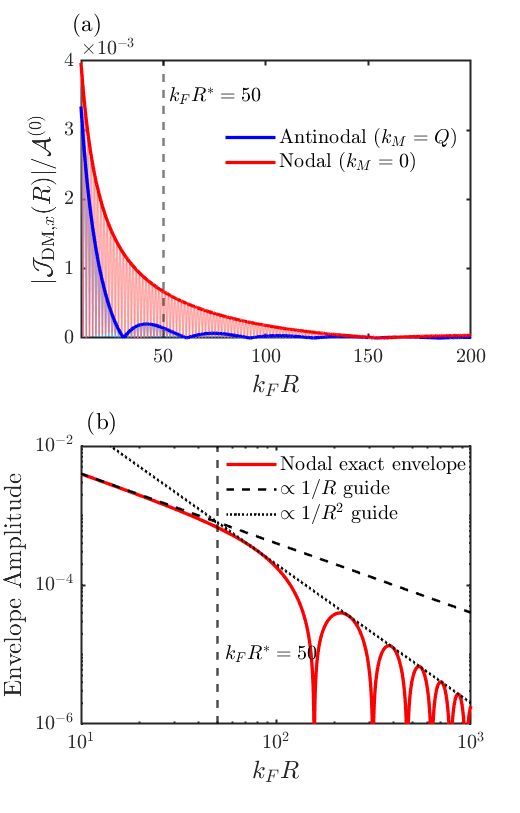}
    \caption{Spatial decay of the absolute in-plane DM interactions in a two-dimensional $p$-wave magnet, demonstrating the anomalous $1/R$ power-law crossover. (a) Linear scale comparison of the spatial interactions. Along the antinodal direction ($k_M = Q$, blue), the interaction decays rapidly, bounded by the standard 2D $1/R^2$ scaling. Conversely, along the nodal line ($k_M = 0$, red), the retention of the Rashba gap yields a massively extended, long-range interaction profile. We use $\mathcal{A}^{(0)}=J_{\text{imp}}^2m/2\pi^2\hbar^2$ as normalization factor. (b) Log-log scale explicitly proving the dimension-reduction effect for the nodal envelope. Inside the intermediate-distance crossover window ($R < R^*$), the regularized envelope (solid red line) tightly follows an effectively 1D-like $1/R$ decay profile (dashed black guide). Once the impurity separation exceeds the characteristic Rashba length $k_F R^* \sim 1/(2k_R)$ (indicated by the vertical dashed lines at $k_F R = 50$), the envelope bends downward, undergoing a mathematical crossover that gracefully reverts to the conventional $1/R^2$ asymptotic tail (dotted black guide). The dimensionless parameters used are $k_R/k_F = 0.01$ and $Q/k_F = 0.05$.}
    \label{fig:fig06}
\end{figure}

\subsection{Anomalous $1/R$ Power-Law Crossover of the In-Plane DM Interaction}
\label{subsec:decay}

The most striking spatial consequence of the nodal geometry appears in the decay profile of the in-plane DM components. In a conventional two-dimensional electron gas, all RKKY interaction channels follow an asymptotic $1/R^2$ spatial decay, governed by the isotropic prefactor $\mathcal{A}(R)$. However, retaining the hybridized Rashba gap reveals an important anomaly along the $p$-wave nodal lines.

By evaluating the expression, Eq.~(\ref{eq:J_DMx_reg}), along the nodal direction where the macroscopic $p$-wave shift vanishes ($k_M = 0$), the momentum splitting is determined purely by the Rashba coupling ($\kappa = k_R$). Setting $\kappa = k_R$ makes the spin-projection factor $k_R/\kappa$ equal to unity, so the in-plane DM term reduces to
\begin{equation}
\mathcal{J}_{\text{DM},x}^{\text{nodal}} \simeq -\mathcal{A}(R) \sin\phi \sin(2k_R R) \sin(2k_F R),
\label{eq:J_DMx_nodal}
\end{equation}
with an analogous expression for $\mathcal{J}_{\text{DM},y}$. Because the RSOC is generically weak in physical materials ($k_R \ll k_F$), it defines an extended, macroscopic characteristic spatial length scale $R^* \sim 1/(2k_R)$. For intermediate impurity separations that are large compared to the Fermi wavelength but smaller than this relativistic length, the envelope function linearizes, $\sin(2k_R R) \approx 2k_R R$. 

Multiplying this linear spatial dependence by the 2D RKKY amplitude $\mathcal{A}(R) \propto 1/R^2$ yields a partial cancellation of $1/R$ decay. This results in an anomalously enhanced, long-range $1/R$ power-law decay given by
\begin{equation}
\mathcal{J}_{\text{DM},x}^{\text{nodal}} \propto -\frac{2k_R}{R} \sin\phi \sin(2k_F R).
\end{equation}
For $R \gtrsim R^*$, the sine envelope bounds the response, and the conventional $1/R^2$ decay is recovered. Thus, the anomalous $1/R$ scaling is not a mathematical divergence but a defined finite intermediate-distance crossover window
\begin{equation}
\frac{1}{Q} \ll R \lesssim R^* \sim \frac{1}{2k_R},
\label{eq:R_crossover}
\end{equation}
which exists whenever $k_R \ll Q$. For the parameters used here ($k_R/k_F = 0.01$), this crucial crossover boundary sits at $k_F R^* \sim 50$.

This dimension-reducing spatial scaling is depicted in Fig.~\ref{fig:fig06}. As shown in the linear-scale comparison of Fig.~\ref{fig:fig06}(a), the antinodal DM interaction (blue curve) decays rapidly, bounded by the standard 2D $1/R^2$ envelope. In contrast, the nodal direction (red curve) exhibits a massively extended interaction range. The proof of this crossover is explicitly demonstrated in the log-log scaling of Fig.~\ref{fig:fig06}(b). Inside the intermediate-distance window, the regularized nodal envelope tightly hugs the dashed $1/R$ power-law guide, mimicking the extended exchange profile of a one-dimensional electron gas. However, once the impurity separation exceeds the characteristic Rashba length scale, marked by the vertical dashed line at $k_F R^* = 50$, the envelope bends downward. Beyond this boundary, the interaction undergoes a graceful mathematical crossover, with its subsequent amplitude peaks reverting to the conventional $1/R^2$ asymptotic tail.

Four physical cutoffs delimit this dramatic enhancement in practice. First, a small angular misalignment $\delta$ from the nodal direction introduces a finite $k_M \approx Q\delta$, meaning the effectively 1D nodal envelope persists only while $Q\delta R \lesssim 1$. Second, the intrinsic RSOC length scale bounds the crossover at $R^* \sim 1/(2k_R)$. Third, finite temperature cuts off the RKKY oscillations beyond the thermal coherence length $\xi_T \sim \hbar v_F / k_B T$. Fourth, elastic disorder in the host introduces a mean free path $\ell$. The disorder-averaged RKKY coupling is exponentially suppressed for $R\gtrsim\ell$, although the typical (sample-specific) coupling retains the clean power law \cite{deGennes1962,Zyuzin1986,Bulaevskii1986,Jagannathan1988,Chesi2010PRB}. The extended $1/R$ nodal response is therefore practically realized within $1/Q \ll R \lesssim \min(R^*, \xi_T,  \ell)$.

Crucially, this dimension-reducing $1/R$ crossover is not limited to $p$-wave systems. As we demonstrate in Appendix \ref{app:fwave}, the $1/R$ window and its slope survive in $f$-wave magnets. Despite the presence of multiple symmetry-enforced nodal lines and a more complex, warped Fermi surface geometry, the anomalous decay holds universally for odd-parity magnets provided the Rashba gap dominates the curvature-induced nodal splitting.

This anomalously enhanced, directionally filtered chiral coupling provides a theoretical mechanism for device engineering. By leveraging this finite 1D-like crossover window, one could theoretically stabilize extended, one-dimensional chains of non-collinear spin spirals or magnetic skyrmions along specific crystallographic axes of unconventional $p$-wave hosts \cite{Schmitt2019}.

\subsection{Material Realizations and Lattice Effects}
\label{subsec:materials}
Candidate materials such as $\mathrm{NiI_2}$ have been experimentally shown to host $p$-wave magnetic order \cite{Song2025Nature}, and metallic $p$-wave magnets, as required for the RKKY mechanism, have also been reported \cite{Yamada2025Nature}. In such realistic materials, the electronic structure exhibits multi-band effects, with multiple bands crossing the Fermi level. The presence of multiple unperturbed Fermi wavevectors, $k_{F,n}$, will inherently result in a superposition of RKKY responses. This introduces multiple characteristic wavelengths, transforming the single-beating profiles described here into more complex, multi-frequency spatial interference patterns. Nevertheless, our central physical conclusions remain fundamentally robust. The anomalous $1/R$ nodal crossover and the parity-driven role reversal, specifically the decoupling of the Ising interaction from the macroscopic beating, do not rely on a specific dispersion relation. Rather, they are rigorously dictated by the odd-parity symmetry of the momentum-dependent exchange field. Because this symmetry is protected by the macroscopic crystalline and magnetic space groups, every individual band pocket inheriting this spin splitting will independently enforce the odd-parity constraint on the magnetic Green's function ($G_z(-\bm{R}) = -G_z(\bm{R})$). More generally, since the parity relations of Sec.~\ref{subsec:momentumGF} follow from the effective time-reversal invariance alone, they hold for the full multi-band Green's function, including its interband contributions. The role-reversed sign structure of Eqs.~(\ref{eq:Jxx}) to (\ref{eq:Jzz}) is holds for any $\mathcal{T}\tau$-symmetric host, with the products of propagators generalized to sums over the internal indices related by $\tau$, and only the wavevectors and amplitudes entering them are material specific. Consequently, while multi-band effects will enrich the short-range oscillatory behavior, the macroscopic parity decoupling and the extended intermediate-distance chiral interactions will robustly survive in realistic $p$-wave hosts.

Furthermore, while our analytical framework employs a parabolic continuum model to capture the physics, our fundamental conclusions remain robust beyond the $k^2$ dispersion limit. A full tight-binding lattice treatment would introduce higher-order band warping, anisotropic Fermi velocities, and van Hove singularities, which can quantitatively modify the short-range RKKY oscillation amplitudes. For a non-circular Fermi surface the stationary-phase points are those where the Fermi-surface normal is parallel to $\bm{R}$, so that $2k_F$ is replaced by the direction-dependent extremal caliper of the Fermi surface, as in the standard treatment of RKKY interactions for non-spherical Fermi surfaces \cite{Roth1966PR}. However, the long-distance asymptotic RKKY exchange ($k_F R \gg 1$) is governed by the poles of the Green's functions at the Fermi surface. Because the odd-parity constraint of the $p$-wave exchange ($M_{-\bm{k}}^p = -M_{\bm{k}}^p$) and the existence of symmetry-enforced nodal lines are protected, the resulting odd parity of the real-space spin propagator ($G_z(-\bm{R}) = -G_z(\bm{R})$) is robust. Consequently, the parity-driven structural decoupling of the Ising channel and the $1/R$ dimension-reducing crossover at the nodes will strictly persist in realistic lattice models.

The spin-flip term $\eta J_{\rm sd}\sigma_x$ of the minimal model of Ref.~\cite{Brekke2024} (Sec.~II.A) can be incorporated in the present formalism. It is even in $\bm{k}$, so on its own it generates no DM interaction (its contributions from the two sectors cancel), and the effective time-reversal operation now exchanges the sectors, $\sigma_y\hat{H}_\eta^{T}(\bm{k})\sigma_y=\hat{H}_{-\eta}(-\bm{k})$, so that the parity-driven sign structure of Eqs.~(\ref{eq:Jxx}) to (\ref{eq:JDMz}) is preserved with the propagators of the two sectors paired in the RKKY trace. At the stationary point the term enters the hybridized denominator on the same footing as the RSOC term, and the regularized splitting of sector $\eta$ becomes $\kappa_\eta^2=k_M^2+(k_R\sin\phi+\eta q_J)^2+k_R^2\cos^2\phi$ with $q_J\simeq mJ_{\rm sd}/(\hbar^2k_F)$. Three consequences follow. For $\lambda=0$ the Ising channel remains free of macroscopic beating, as in Eq.~(\ref{eq:J_zz_simp}), because the spin-flip contribution combines with $G_z^2$ into a non-beating term ($k_M^2+q_J^2=\kappa^2$), while the in-plane channels beat with different contrasts along the spin axes parallel and perpendicular to the spin-flip field. Along a nodal line the $1/R$ regime of Eq.~(\ref{eq:R_crossover}) survives with the same slope, but its extent depends on the orientation of the in-plane DM component relative to the spin-flip axis. For the component perpendicular to that axis the two gaps add in quadrature and $R^*\sim 1/(2k_R)$ is replaced by $1/(2\sqrt{k_R^2+q_J^2})$, whereas for the component along it the leading-order nodal result of Eq.~(\ref{eq:J_DMx_nodal}) is unchanged. The anomalous regime is therefore controlled by the net nodal gap rather than by the Rashba gap alone.

Two further remarks concern the robustness of the host. First, non-magnetic disorder does not spoil the parity structure. A scalar potential $V(\bm{r})\sigma_0$ preserves the effective time-reversal invariance, so the relation $\hat{G}(\bm{R}_1,\bm{R}_2)=\sigma_y\hat{G}^{T}(\bm{R}_2,\bm{R}_1)\sigma_y$, and with it the role-reversed sign structure of Sec.~\ref{subsec:roles}, holds for every disorder configuration. Disorder averaging therefore modifies the amplitudes, through the mean free path discussed in Sec.~\ref{subsec:decay}, but not the sign structure that assigns the macroscopic beating to the in-plane channels (in the self-consistent Born approximation the impurity self-energy is proportional to $\sigma_0$, because the $\bm{k}$-odd spin components of the local propagator vanish) \cite{Chesi2010PRB}. Second, the host order is treated here, as in all RKKY calculations, as a rigid mean-field background. This is justified for two dilute adatoms, whose back-action on the itinerant spin density is a local, Friedel-like perturbation \cite{Sukhachov_Friedel2024} that does not feed back on the exchange at order $J_{\text{imp}}^2$. Whether a finite density of magnetic impurities could renormalize or destabilize the comparatively soft non-collinear $p$-wave order is an interesting question that lies beyond the two-impurity problem studied here.

\begin{figure*}[t]
    \centering
    \includegraphics[width=\textwidth]{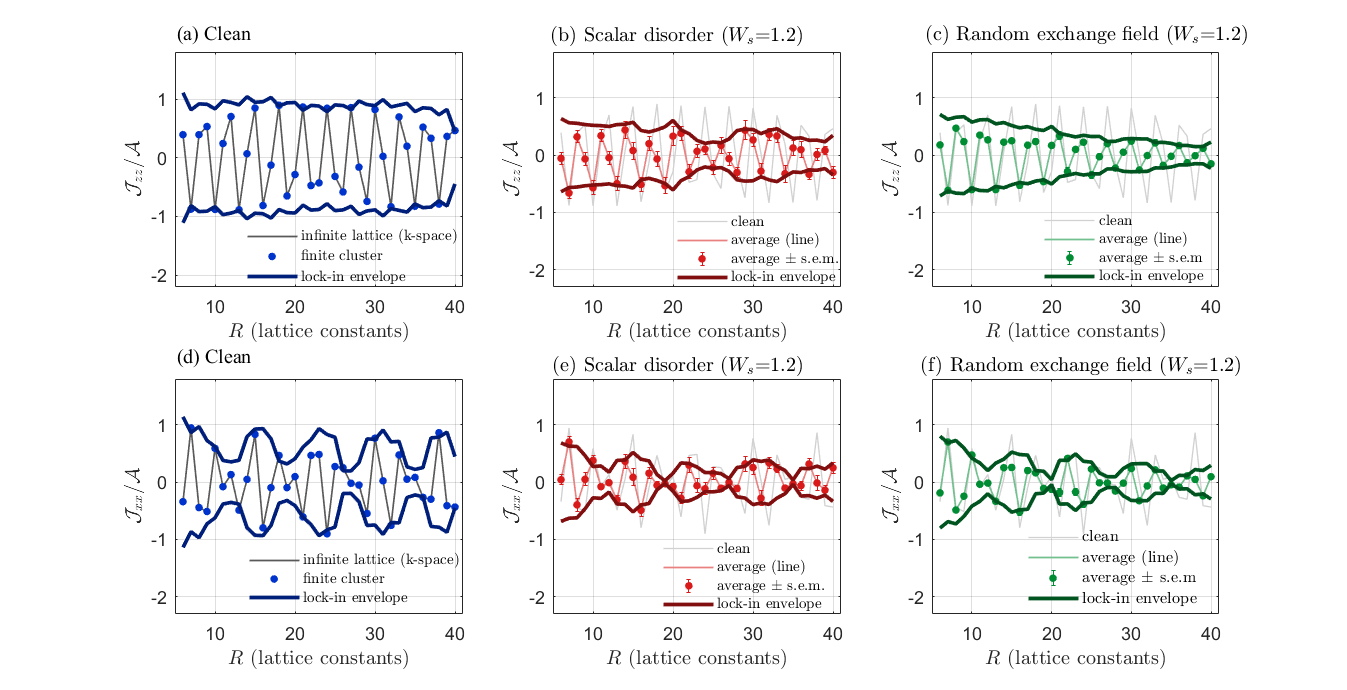}
     \caption{Numerical tight-binding verification of the parity-driven RKKY decoupling in the presence of disorder. The panels show the spatial dependence of the out-of-plane Ising component $J_{zz}$ [(a)–(c)] and the in-plane Heisenberg component $J_{xx}$ [(d)–(f)], normalized by the standard isotropic decay amplitude $\mathcal{A}(R)$. The calculations are performed on a two-dimensional square lattice with complex absorbing boundaries to eliminate finite-size reflections. Columns correspond to the clean limit [(a), (d)], random scalar potential disorder [(b), (e)], and random magnetic exchange disorder [(c), (f)]. Thin lines connecting markers represent the disorder-averaged RKKY signal, with error bars indicating the standard error of the mean (s.e.m.) over 40 configurations. Thick solid lines trace the extracted spatial envelopes. While both scalar and magnetic disorder introduce an exponential damping due to a finite mean free path, the fundamental structural decoupling is preserved: the Ising channel escapes the macroscopic modulation (flat envelope), whereas the in-plane Heisenberg channel retains the robust $p$-wave beating. The dimensionless lattice parameters perfectly match the continuum limits used in previous sections: $Q/k_F = 0.20$ and $k_R/k_F = 0.05$, with a matched disorder strength variance of $W_s^2/12$ for both impurity types.}
    \label{fig:TB_verification}
\end{figure*}

\subsection{Numerical Tight-Binding Verification and Robustness to Disorder}
\label{sec:numerical_disorder}

To validate our analytical findings and verify the robustness of the parity-driven decoupling against disorder, we evaluate the RKKY interaction using a real-space tight-binding model. This numerical framework serves a dual purpose: it confirms that our conclusions persist beyond the parabolic continuum limit (incorporating lattice effects), and it addresses the specific physical scenario of soft, non-collinear $p$-wave magnets, where random magnetic exchange disorder is an expected ingredient.

The lattice is modeled on a two-dimensional square lattice driven by the tight-binding counterpart of Eq. (\ref{eq:1}). In momentum space, the discrete Hamiltonian reads 

\begin{align}
    \hat{H}(\mathbf{k}) = & [4t - 2t(\cos k_x a + \cos k_y a) ]\sigma_0 \nonumber \\
    &- 2t_p[\cos(\beta)\sin k_x a + \sin(\beta)\sin k_y a]\sigma_z \nonumber \\
    & + 2\alpha[\sin(k_y a)\sigma_x - \sin(k_x a)\sigma_y],
\end{align}

where $a$ is the lattice constant, $t = \hbar^2/(2ma^2)$ is the hopping parameter, and the effective $p$-wave and Rashba couplings are given by $2t_p a = J/k_F$ and $2\alpha a = \lambda$, respectively. To eliminate finite-size edge reflections and wrap-around effects that plague periodic clusters, we evaluate the exact Matsubara Green's functions, $\hat{G}(\bm{R}, i\omega_n) = [i\omega_n\sigma_0 - \hat{H} + i\hat{\Gamma}]^{-1}$, on an open cluster terminated by complex absorbing layers modeled by a spatially varying self-energy term $i\hat{\Gamma}$. We align the impurity bond along the $x$-axis ($\phi = 0$) and set the $p$-wave polarization to $\beta = 0$. The hopping parameters are chosen such that the low-energy effective scales precisely match the parameters utilized in the continuum plots: $Q/k_F = 0.20$ and $k_R/k_F = 0.05$.

We investigate three distinct scenarios: (i) a clean lattice, (ii) random scalar potential disorder modeled by an Anderson-type on-site potential $V \in [-W_s/2, W_s/2]$, and (iii) a random magnetic exchange field $\bm{h} \cdot \bm{\sigma}$, where the local field $\bm{h}$ has an isotropic random orientation on the Bloch sphere and a uniform random magnitude $|\bm{h}| \in [0, W_s/2]$. To have a direct comparison between the two disorder types, the maximum amplitudes are constrained to yield identical second moments, $\langle V^2 \rangle = \langle |\bm{h}|^2 \rangle = W_s^2/12$. This equivalence guarantees that both disorder potentials generate the same Born scattering rates and identical mean free paths.

The numerical results are presented in Fig.~\ref{fig:TB_verification}. In the clean limit [Figs.~\ref{fig:TB_verification}(a) and \ref{fig:TB_verification}(d)], the lattice calculation reproduces the central analytical prediction of Sec.~\ref{subsec:roles} (gray lines indicate the momentum-space integration for the infinite clean lattice). To extract these spatial envelopes from the numerical data, we apply a spatial lock-in demodulation procedure: the oscillatory signal is multiplied by $e^{-i q_c R}$ (where $q_c$ is the dominant numerical oscillation frequency) and subsequently smoothed using a moving average. The out-of-plane Ising component $J_{zz}$ exhibits standard oscillations without beating and modulation [Fig.~\ref{fig:TB_verification}(a)]. In contrast, the in-plane Heisenberg component $J_{xx}$ is modulated, exhibiting a spatial beating envelope dictated by the hybridized wavevector $\kappa$ [Fig.~\ref{fig:TB_verification}(d)]. 

Crucially, the introduction of both scalar [Figs.~\ref{fig:TB_verification}(b) and \ref{fig:TB_verification}(e)] and random magnetic [Figs.~\ref{fig:TB_verification}(c) and \ref{fig:TB_verification}(f)] disorder leaves this parity-driven decoupling intact. In the disordered panels, the discrete markers denote the ensemble-averaged RKKY signal, while the error bars represent the standard error of the mean (s.e.m.) computed over 40 independent configurations. As expected, disorder suppresses the overall RKKY amplitude exponentially over a characteristic mean free path $\ell$. However, factoring out this exponential damping reveals that the structural spatial symmetries are highly resilient. The random magnetic exchange field---despite locally breaking time-reversal symmetry and fluctuating in all three spatial dimensions---does not mix the decoupled RKKY channels nor destroy the $p$-wave beating.

This striking robustness is intimately tied to the defining symmetries of the $p$-wave state itself. In the self-consistent Born approximation, the local magnetic self-energy scales as $\Sigma_{\text{mag}} \propto \sum_{\bm{k}} \langle (\bm{h}\cdot\bm{\sigma}) \hat{G}(\bm{k}) (\bm{h}\cdot\bm{\sigma}) \rangle$. Because the $p$-wave exchange field dictates that the spin-dependent vector components of the Green's function $\bm{G}(\bm{k})$ are odd in momentum space (Sec.~\ref{subsec:momentumGF}), their integral over the Brillouin zone vanishes. Consequently, the local magnetic scattering effectively reduces to a spin-independent scalar self-energy, $\Sigma_{\text{mag}} \propto \langle |\bm{h}|^2 \rangle \sum_{\bm{k}} G_0(\bm{k})\sigma_0$. Therefore, random magnetic disorder in a $p$-wave magnet behaves similar to scalar disorder.

\section{CONCLUSION}

In summary, we have developed a comprehensive analytical framework for the RKKY interaction in two-dimensional odd-parity magnets subjected to RSOC. By deriving closed-form expressions for the real-space Green's functions, our study uncovers how the defining odd-parity symmetry of the exchange field fundamentally reshapes the landscape of indirect magnetic exchange, establishing a phenomenology distinct from both conventional ferromagnets and even-parity $d$-wave altermagnets. By expanding our framework from $p$-wave to $f$-wave systems, we proved that the parity-driven decoupling and the anomalous nodal scaling manifest universally across odd-parity magnetic structures.

Our principal findings reveal a profound parity-driven ``role reversal'' in the spatial magnetic responses. Specifically, the odd-parity momentum shift structurally decouples the out-of-plane Ising interaction ($\mathcal{J}_{zz}$) from the macroscopic $p$-wave modulation. Consequently, $\mathcal{J}_{zz}$ sheds the anisotropic macroscopic beating entirely, oscillating exclusively at the shifted Fermi-surface radial wavevector $2k_F$ in weak-RSOC regime. In stark contrast, the in-plane Heisenberg components ($\mathcal{J}_{xx}, \mathcal{J}_{yy}$) exhibit robust, highly directional spatial beating enveloped by the anisotropic $p$-wave wavevector.

Beyond collinear exchange, we identified a rich, multi-component non-collinear response. Away from the nodal lines, the macroscopic $p$-wave momentum shift generates a massive, nonrelativistic out-of-plane chiral component ($\mathcal{J}_{\text{DM},z}$). Crucially, the interplay between the unconventional magnetic order and relativistic spin-momentum locking yields a striking spatial anomaly along the $p$-wave nodal lines. Because the macroscopic momentum shift vanishes at the nodes, the intrinsic Rashba gap prevents mathematical divergences and instead leads to an anomalous dimensional reduction. We demonstrated that the in-plane DM interaction components crossover from the conventional 2D $1/R^2$ scaling into a massively enhanced, 1D-like $1/R$ power-law decay over a macroscopic intermediate-distance window, $1/Q \ll R \lesssim 1/(2k_R)$, before ultimately reverting to the $1/R^2$ asymptotic tail. 

Furthermore, our numerical tight-binding calculations demonstrate that this striking parity-driven decoupling is highly resilient. It survives not only standard scalar impurities but also the random magnetic exchange disorder that is characteristically expected in soft non-collinear magnetic phases. 

From a fundamental perspective, these results provide a physical and mathematical blueprint for two-body interactions in odd-parity magnetic ground states. Crucially, from a device-engineering standpoint, this mechanism offers concrete advantages over existing spintronic approaches based on conventional ferromagnets. In standard 2D platforms, the Heisenberg, Ising, and DM exchange channels are highly entangled and universally decay as $1/R^2$, often complicating the isolation of chiral interactions. In contrast, the structural insulation of the Ising channel and the directionally filtered $1/R$ nodal DM interaction identified here provide a unique toolkit to avoid such channel entanglement. The same decoupling also offers a concrete spectroscopic advantage for probing the band structure. In a spin-resolved measurement of adatom pairs \cite{Zhou2010}, the period of the out-of-plane channel yields the Fermi wavevector $k_F$ alone, while the beating period $\pi/\kappa$ of the in-plane channels, recorded as a function of the bond orientation $\phi$, maps out $Q\cos(\phi-\beta)$ and hence both the magnitude and the direction of the $p$-wave splitting vector. In a conventional Rashba electron gas all channels beat together at the single relativistic scale $2k_R$, and in an even-parity host the two roles are interchanged, so that the assignment of the beating to the in-plane rather than to the out-of-plane channel is itself a fingerprint of odd-parity order. These features identify $p$-wave magnets as exceptionally promising platforms, offering highly tunable, orientation-driven mechanisms to stabilize low-dimensional non-collinear spin textures.

\section{ACKNOWLEDGMENTS}
M. S. greatly thanks R. Beiranvand for helpful discussions. 
T. F. was supported by the Research Council of Norway through its Centers of Excellence funding scheme, Project No. 353919 and Project No. 361800 ``QTransMag''. We acknowledge the use of Gemini 3.1 Pro for English language editing and stylistic refinement of this manuscript.
 
\section{DATA AVAILABILITY}
The data that support the findings of this article are not
publicly available. The data are available from the authors
upon reasonable request.

\bibliography{references} 

\widetext
\appendix
\section{REAL-SPACE GREEN'S FUNCTIONS}
\label{Appendix A}
In this Appendix, we present the detailed mathematical derivations of the real-space Green's functions $G_n(\bm{R}, \mathcal{E})$ for $n \in \{0, x, y, z\}$ used in Section II.C. To capture the relativistic spin-orbit physics globally, we directly evaluate the spatial integrals by retaining the full Rashba term in the Green's function denominator. The real-space transformation is defined as,
\begin{equation}
    G_n(\bm{R}, \mathcal{E}) = \int \frac{d^2k}{(2\pi)^2} e^{i\bm{k}\cdot\bm{R}} G_n(\bm{k}, \mathcal{E}).
\end{equation}

To evaluate these integrals analytically, we express the momentum vector in polar coordinates $\bm{k} = (k\cos\theta, k\sin\theta)$ and the spatial separation as $\bm{R} = (R\cos\phi, R\sin\phi)$. For large impurity separations ($k_F R \gg 1$), the highly oscillatory angular integration is dominated by two stationary phase points at $\theta_1 = \phi$ (forward) and $\theta_2 = \phi + \pi$ (backward). Applying the standard stationary phase approximation for an integral containing a generic slowly varying amplitude function $F(\theta)$ (which in our case represents the non-exponential angular dependence of the Green's function $G_n$), the angular integral yields the asymptotic Hankel functions of the first and second kind, respectively,
\begin{eqnarray}
    \int_0^{2\pi} d\theta e^{ikR\cos(\theta-\phi)} F(\theta) &\approx& \sqrt{\frac{2\pi}{kR}} e^{i(kR-\pi/4)} F(\phi) + \sqrt{\frac{2\pi}{kR}} e^{-i(kR-\pi/4)} F(\phi+\pi) \nonumber \\
    &\approx& \pi H_0^{(1)}(kR) F(\phi) + \pi H_0^{(2)}(kR) F(\phi+\pi).
    \label{eq:A2}
\end{eqnarray}
Substituting this into the 2D Fourier transform leaves a 1D radial integral evaluated from $0$ to $\infty$
\begin{equation}
    G_n(\bm{R}, \mathcal{E}) \approx \frac{1}{4\pi} \int_0^\infty k dk \Big[ H_0^{(1)}(kR) G_n(k, \phi, \mathcal{E}) + H_0^{(2)}(kR) G_n(k, \phi+\pi, \mathcal{E}) \Big].
\end{equation}
We can elegantly simplify this by using the analytic continuation $H_0^{(2)}(kR) = -H_0^{(1)}(-kR)$. Substituting $k \to -k$ in the second term maps its integration bounds to $(-\infty, 0)$. Because a momentum vector with a negative radius at angle $\phi+\pi$ is physically identical to a positive radius at angle $\phi$, we have $G_n(-k, \phi+\pi, \mathcal{E}) = G_n(k, \phi, \mathcal{E})$ for all spin components. The two terms thus merge seamlessly into a single integral over the entire real axis,
\begin{equation}
    G_n(\bm{R}, \mathcal{E}) \approx \frac{1}{4\pi} \int_{-\infty}^\infty k dk \, H_0^{(1)}(kR) G_n(k, \phi, \mathcal{E}).
\end{equation}

At the stationary angle $\theta = \phi$, the unified denominator shared by all Green's function components fundamentally simplifies to,
\begin{equation}
    \mathcal{D}(k, \phi) = A_{\bm{k}}^2 - (\Delta_p k \cos(\phi-\beta))^2 - \lambda^2 k^2 = A_{\bm{k}}^2 - \left(\frac{\hbar^2}{m} \kappa k\right)^2,
\end{equation}
where $A_{\bm{k}} = \mathcal{E} + i\eta - \frac{\hbar^2 k^2}{2m}$.  The hybridized split poles are located at $\mathcal{D}(k, \phi) = 0$, giving $A_{\bm{k}} = \pm \frac{\hbar^2}{m} \kappa k$. Expanding around the hybridized center wavevector $k_0$, these poles reside at $k_\pm = k_0 \pm \kappa$ on the real axis. By factoring the denominator as $\mathcal{D}(k, \phi) \propto (k^2 - k_+^2)(k^2 - k_-^2)$, we apply the Cauchy residue theorem to perform the $k$-integration for each respective numerator.

For the scalar component $G_0$ and the out-of-plane $p$-wave component $G_z$, the momentum-space numerators at $\theta = \phi$ are $N_0 = A_{\bm{k}}$ and $N_z = -\frac{\hbar^2}{m} k_M k$, respectively. Evaluating the exponential residues at $k_\pm$ directly yields the closed-form split-phase representations,
\begin{eqnarray}
    G_0(\bm{R}, \mathcal{E}) &=& -\frac{im}{4\hbar^2} \Big[ H_0^{(1)}(k_+ R) + H_0^{(1)}(k_- R) \Big], \label{eq:G0_app} \\
    G_z(\bm{R}, \mathcal{E}) &=& -\frac{im}{4\hbar^2} \frac{k_M}{\kappa} \Big[ H_0^{(1)}(k_+ R) - H_0^{(1)}(k_- R) \Big]. \label{eq:Gz_app}
\end{eqnarray}
Notice that the hybridization with the RSOC endows $G_z$ with the dimensionless spin-projection factor $k_M/\kappa$.

We next evaluate the off-diagonal elements $G_x(\bm{R}, \mathcal{E})$ and $G_y(\bm{R}, \mathcal{E})$, whose numerators evaluate at the stationary point to $N_x = \lambda k \sin\phi$ and $N_y = -\lambda k \cos\phi$. Because these numerators scale linearly with $k$, the extra factor of $k$ in the radial integrand acts as a spatial derivative with respect to $R$, which raises the order of the Hankel function, mapping $H_0^{(1)}(k_\pm R) \to H_1^{(1)}(k_\pm R)$, and pulls down the respective split momenta $k_\pm$. Applying the identical residue procedure yields the closed-form expressions,
\begin{align}
    G_x(\bm{R}, \mathcal{E}) &= \frac{ m \sin\phi}{4\hbar^2}\frac{k_R}{\kappa}\frac{1}{k_0} \left[ k_+ H_1^{(1)}(k_+R) - k_- H_1^{(1)}(k_-R) \right], \label{eq:Gx_app} \\
    G_y(\bm{R}, \mathcal{E}) &=\frac{- m \cos\phi}{4\hbar^2}\frac{k_R}{\kappa}\frac{1}{k_0} \left[ k_+ H_1^{(1)}(k_+R) - k_- H_1^{(1)}(k_-R) \right]. \label{eq:Gy_app}
\end{align}

\section{THE RKKY EXCHANGE TENSOR COMPONENTS}
\label{Appendix B}

In this Appendix, we provide the step-by-step derivation of the RKKY exchange tensor components, starting from the general zero-temperature RKKY Hamiltonian obtained via second-order perturbation theory,
\begin{equation}
    \mathcal{H}_{\text{RKKY}} = -\frac{J_{\text{imp}}^2}{\pi} \text{Im} \int_{-\infty}^{\mathcal{E}_F} d\mathcal{E} \, \text{Tr} \Big[ (\bm{S}_1 \cdot \bm{\sigma}) \hat{G}(\bm{R}) (\bm{S}_2 \cdot \bm{\sigma}) \hat{G}(-\bm{R}) \Big]. \label{eq:15}
\end{equation}

The retarded Green's function in real space is decomposed into its scalar and vector Pauli components as $\hat{G}(\pm\bm{R}) = G_0(\pm\bm{R})\sigma_0 + \bm{G}(\pm\bm{R})\cdot \bm{\sigma}$. As established in the main text, the odd-parity symmetry of the $p$-wave magnetic order dictates that the scalar component is even under spatial inversion, while the vector components are odd. Therefore, we have $G_0(-\bm{R}) = G_0(\bm{R})$ and $\bm{G}(-\bm{R}) = -\bm{G}(\bm{R})$. 

Substituting these parity relations into the trace, we expand the product of the Pauli matrices,
\begin{equation*}
    \text{Tr} \Big[ (\bm{S}_1 \cdot \bm{\sigma}) (G_0 \sigma_0 + \bm{G} \cdot \bm{\sigma}) (\bm{S}_2 \cdot \bm{\sigma}) (G_0 \sigma_0 - \bm{G} \cdot \bm{\sigma}) \Big].
\end{equation*}
Evaluating this requires the standard Pauli matrix trace identities, $\text{Tr}[\sigma_i \sigma_j] = 2\delta_{ij}$, $\text{Tr}[\sigma_i \sigma_j \sigma_k] = 2i\epsilon_{ijk}$, and $\text{Tr}[\sigma_i \sigma_j \sigma_k \sigma_l] = 2(\delta_{ij}\delta_{kl} - \delta_{ik}\delta_{jl} + \delta_{il}\delta_{jk})$. Grouping the scalar and cross-product terms, the trace evaluates to,
\begin{align*}
    \text{Tr}[...] &= 2 \Big(G_0^2 + |\bm{G}|^2\Big) (\bm{S}_1 \cdot \bm{S}_2) - 4 (\bm{S}_1 \cdot \bm{G})(\bm{S}_2 \cdot \bm{G}) \\
    &\quad - 4i G_0 \bm{G} \cdot (\bm{S}_1 \times \bm{S}_2),
\end{align*}
where $|\bm{G}|^2 = G_x^2 + G_y^2 + G_z^2$, and the spatial argument $\bm{R}$ is implied for all Green's functions. The antisymmetric contribution carries an overall minus sign, $-4iG_0\bm{G}\cdot(\bm{S}_1\times\bm{S}_2)$. The corresponding global sign of the DM vector is a chirality convention fixed by the labeling of the two impurities (equivalently the orientation $\bm{R}=\bm{R}_2-\bm{R}_1$) and does not affect the magnitudes, beating envelopes, or power-law decays reported in this work. 

Expanding the spin dot products in Cartesian coordinates gives $(\bm{S}_1 \cdot \bm{S}_2) = S_1^x S_2^x + S_1^y S_2^y + S_1^z S_2^z$ and $(\bm{S}_1 \cdot \bm{G})(\bm{S}_2 \cdot \bm{G}) = \sum_{i,j} G_i G_j S_1^i S_2^j$. Substituting this expanded trace back into Eq.~(\ref{eq:15}) allows us to map the Hamiltonian directly onto the general tensorial form,
\begin{equation}
    \mathcal{H}_{\text{RKKY}} = \sum_{\ell=x,y,z}\mathcal{J}_{\ell\ell}(\bm{R})S_1^\ell S_2^\ell
    +\sum_{\ell<m}\mathcal{J}_{\ell m}(\bm{R})\left[S_1^\ell S_2^m+S_1^m S_2^\ell\right]
    +\bm{\mathcal{J}}_{\text{DM}}(\bm{R})\cdot(\bm{S}_1\times\bm{S}_2). \label{eq:16}
\end{equation}
The second sum is restricted to unordered pairs $\ell<m$, so the off-diagonal symmetric terms are not double counted. The overall factor of two multiplying $(G_0^2+|\bm{G}|^2)$ in the trace is intrinsic to the Pauli algebra and is \emph{not} removed by this restriction. It must be retained in the diagonal coefficients below [Eqs.~(\ref{eq:17_app}) to~(\ref{eq:19_app})]. The resulting interactions fall into three groups, which we derive in turn.

The Heisenberg and Ising interactions follow by extracting the coefficients for $S_1^x S_2^x$, $S_1^y S_2^y$, and $S_1^z S_2^z$ from the expanded trace, which gives the diagonal exchange interactions,
    \begin{align}
        \mathcal{J}_{xx}(\bm{R}) &= -\frac{2J_{\text{imp}}^2}{\pi} \text{Im} \int_{-\infty}^{\mathcal{E}_F} d\mathcal{E} \Big[ G_0^2 - G_x^2 + G_y^2 + G_z^2 \Big], \label{eq:17_app} \\
        \mathcal{J}_{yy}(\bm{R}) &= -\frac{2J_{\text{imp}}^2}{\pi} \text{Im} \int_{-\infty}^{\mathcal{E}_F} d\mathcal{E} \Big[ G_0^2 + G_x^2 - G_y^2 + G_z^2 \Big], \label{eq:18_app} \\
        \mathcal{J}_{zz}(\bm{R}) &= -\frac{2J_{\text{imp}}^2}{\pi} \text{Im} \int_{-\infty}^{\mathcal{E}_F} d\mathcal{E} \Big[ G_0^2 + G_x^2 + G_y^2 - G_z^2 \Big]. \label{eq:19_app}
    \end{align}

The symmetric off-diagonal interactions arise from the cross terms in the trace expansion, such as $-4 G_x G_y (S_1^x S_2^y + S_1^y S_2^x)$, which map directly onto the off-diagonal tensor components $\mathcal{J}_{\ell m}$ for $\ell \neq m$. Multiplying by the $-\frac{J_{\text{imp}}^2}{\pi}$ prefactor cancels the negative sign and yields,
    \begin{align}
        \mathcal{J}_{xy}(\bm{R}) &= \frac{4 J_{\text{imp}}^2}{\pi} \text{Im} \int_{-\infty}^{\mathcal{E}_F} d\mathcal{E} \Big[ G_x G_y \Big], \label{eq:20_app} \\
        \mathcal{J}_{yz}(\bm{R}) &= \frac{4 J_{\text{imp}}^2}{\pi} \text{Im} \int_{-\infty}^{\mathcal{E}_F} d\mathcal{E} \Big[ G_y G_z \Big], \label{eq:21_app} \\
        \mathcal{J}_{xz}(\bm{R}) &= \frac{4 J_{\text{imp}}^2}{\pi} \text{Im} \int_{-\infty}^{\mathcal{E}_F} d\mathcal{E} \Big[ G_x G_z \Big]. \label{eq:22_app}
    \end{align}

Finally, the DM vector, which governs non-collinear spin twisting, follows from the antisymmetric part of the trace, $-4i G_0 \bm{G} \cdot (\bm{S}_1 \times \bm{S}_2)$. Substituting this into the RKKY integral [Eq.~(\ref{eq:15})] and equating it to the DM term in Eq.~(\ref{eq:16}) with the sign convention noted above, we isolate the Cartesian components of the DM vector,
    \begin{align}
        \mathcal{J}_{\text{DM},x}(\bm{R}) &= \frac{4 J_{\text{imp}}^2}{\pi} \text{Im} \int_{-\infty}^{\mathcal{E}_F} d\mathcal{E} \Big[ i G_0 G_x \Big], \label{eq:23_app} \\
        \mathcal{J}_{\text{DM},y}(\bm{R}) &= \frac{4 J_{\text{imp}}^2}{\pi} \text{Im} \int_{-\infty}^{\mathcal{E}_F} d\mathcal{E} \Big[ i G_0 G_y \Big], \label{eq:24_app} \\
        \mathcal{J}_{\text{DM},z}(\bm{R}) &= \frac{4 J_{\text{imp}}^2}{\pi} \text{Im} \int_{-\infty}^{\mathcal{E}_F} d\mathcal{E} \Big[ i G_0 G_z \Big]. \label{eq:25_app}
    \end{align}

These tensor expressions establish the framework used to compute the analytically closed-form asymptotic interactions in the main text.

\section{THE ASYMPTOTIC RKKY COMPONENTS}
\label{Appendix C}

In this Appendix, we provide the step-by-step derivation of the long-range asymptotic expressions for the RKKY interaction tensor components. To capture the full spatial and spin structure induced by the relativistic spin-orbit coupling, including intermediate phase shifts, we start from the integral forms of the exchange tensor derived in the main text and evaluate them in the asymptotic limit ($k_F R \gg 1$).

\subsection{Derivation of the Heisenberg and Ising Interactions}

To evaluate the RKKY integrals, we substitute the large-argument asymptotic form of the Hankel functions, $H_1^{(1)}(k_\pm R) \approx -i H_0^{(1)}(k_\pm R) \equiv -i H_\pm$, and write $k_\pm = k_0 \pm \kappa$. This yields
\begin{equation}
k_+ H_+ - k_- H_- = k_0(H_+ - H_-) + \kappa(H_+ + H_-).
\end{equation}

When squaring the Green's functions, we utilize the asymptotic Hankel product approximations, $H_\pm \approx \sqrt{2/(\pi k_0 R)} e^{i(k_\pm R - \pi/4)}$, in which the slowly varying amplitude is evaluated at the shared momentum $k_0$ while the split momenta $k_\pm$ are retained in the rapidly oscillating phases
\begin{align}
(H_+ + H_-)^2 &\approx -\frac{4i}{\pi k_0 R} e^{2ik_0 R} \big[1 + \cos(2\kappa R)\big], \label{eq:H_sum} \\
(H_+ - H_-)^2 &\approx \frac{4i}{\pi k_0 R} e^{2ik_0 R} \big[1 - \cos(2\kappa R)\big], \label{eq:H_diff} \\
H_+^2 - H_-^2 &\approx \frac{4}{\pi k_0 R} e^{2ik_0 R} \sin(2\kappa R). \label{eq:H_cross}
\end{align}

Defining the common prefactor $\mathcal{P}(k_0) = \frac{i m^2}{4\pi\hbar^4 k_0 R} e^{2ik_0 R}$, the squared Green's function components rigorously evaluate to
\begin{align}
G_0^2 &\approx \mathcal{P}(k_0) \big[ 1 + \cos(2\kappa R) \big], \\
G_z^2 &\approx -\mathcal{P}(k_0) \frac{k_M^2}{\kappa^2} \big[ 1 - \cos(2\kappa R) \big], \\
G_{x,y}^2 &\approx -\mathcal{P}(k_0) \frac{k_R^2}{\kappa^2} \left\{\begin{matrix} \sin^2\phi \\ \cos^2\phi \end{matrix}\right\} \left[ 1 - \cos(2\kappa R) - \frac{\kappa^2}{k_0^2}\big(1 + \cos(2\kappa R)\big) - \frac{2i\kappa}{k_0} \sin(2\kappa R) \right].
\end{align}
Dropping the $\kappa/k_0$ terms in $G_{x,y}^2$ recovers the leading-order expansion, but retaining them is necessary to capture the higher-order phase shifts. The diagonal exchange interactions require specific sums and differences of these squared terms, integrated over energy up to the Fermi level, $\mathcal{E}_F$. Changing the integration variable via $d\mathcal{E} = \frac{\hbar^2}{m}k_0 dk_0$, the interactions take the form
\begin{equation}
\mathcal{J}_{\ell\ell}(\bm{R}) = -\frac{2J_{\text{imp}}^2}{\pi} \text{Im} \int_0^{k_F} \frac{\hbar^2 k_0}{m} dk_0 \sum G_i^2.
\end{equation}

Inserting $\mathcal{P}(k_0)$, the momentum integrals reduce to a standard form. Factoring out the isotropic 2D RKKY spatial amplitude $\mathcal{A}(R) = \frac{J_{\text{imp}}^2 m}{2\pi^2\hbar^2 R^2}$, we utilize the integration mappings evaluated at the Fermi surface
\begin{align}
\text{Im} \left[ i \int_0^{k_F} e^{2ik_0 R} dk_0 \right] &\approx \frac{\sin(2k_F R)}{2R}, \label{Eq.68}\\
\text{Im} \left[ i \int_0^{k_F} \frac{e^{2ik_0 R}}{k_0} dk_0 \right] &\approx \frac{\sin(2k_F R)}{2R k_F}, \label{Eq.69}\\
\text{Im} \left[ i \int_0^{k_F} i\frac{e^{2ik_0 R}}{k_0} dk_0 \right] &= \text{Im} \left[ - \int_0^{k_F} \frac{e^{2ik_0 R}}{k_0} dk_0 \right] \approx \frac{\cos(2k_F R)}{2R k_F}.\label{Eq.70}
\end{align}
Applying this integration rule systematically maps the real phase components to $\sin(2k_F R)$ and the imaginary components to $\cos(2k_F R)$, generating sub-leading corrections proportional to $1/k_F$. 

By grouping the terms strictly according to their $\mathcal{J}_{xx}$ ($G_0^2 - G_x^2 + G_y^2 + G_z^2$), $\mathcal{J}_{yy}$ ($G_0^2 + G_x^2 - G_y^2 + G_z^2$), and $\mathcal{J}_{zz}$ ($G_0^2 + G_x^2 + G_y^2 - G_z^2$) trace combinations and utilizing $\kappa^2 = k_M^2 + k_R^2$, we arrive at the full analytically closed-form diagonal interactions

\begin{align}
\mathcal{J}_{xx}(\bm{R}) &\simeq -\mathcal{A}(R) \left[ \left( \frac{k_R^2 \sin^2\phi}{\kappa^2} + \frac{k_R^2 \cos(2\phi)}{2k_F^2} \right) + \left( \frac{k_M^2 + k_R^2 \cos^2\phi}{\kappa^2} + \frac{k_R^2 \cos(2\phi)}{2k_F^2} \right) \cos(2\kappa R) \right] \sin(2k_F R) \nonumber \\
&\quad - \mathcal{A}(R) \frac{k_R^2 \cos(2\phi)}{\kappa k_F} \sin(2\kappa R) \cos(2k_F R), \label{eq:Jxx_full}
\end{align}

\begin{align}
\mathcal{J}_{yy}(\bm{R}) &\simeq -\mathcal{A}(R) \left[ \left( \frac{k_R^2 \cos^2\phi}{\kappa^2} - \frac{k_R^2 \cos(2\phi)}{2k_F^2} \right) + \left( \frac{k_M^2 + k_R^2 \sin^2\phi}{\kappa^2} - \frac{k_R^2 \cos(2\phi)}{2k_F^2} \right) \cos(2\kappa R) \right] \sin(2k_F R) \nonumber \\
&\quad + \mathcal{A}(R) \frac{k_R^2 \cos(2\phi)}{\kappa k_F} \sin(2\kappa R) \cos(2k_F R), \label{eq:Jyy_full}
\end{align}

\begin{align}
\mathcal{J}_{zz}(\bm{R}) &\simeq -\mathcal{A}(R) \left[ \left( \frac{k_M^2}{\kappa^2} + \frac{k_R^2}{2k_F^2} \right) + \left( \frac{k_R^2}{\kappa^2} + \frac{k_R^2}{2k_F^2} \right) \cos(2\kappa R) \right] \sin(2k_F R) \nonumber \\
&\quad - \mathcal{A}(R) \frac{k_R^2}{\kappa k_F} \sin(2\kappa R) \cos(2k_F R). \label{eq:Jzz_full}
\end{align}

These expressions reveal two physical hallmarks of the Rashba interaction. First, it rigorously breaks the in-plane spatial isotropy ($\mathcal{J}_{xx} \neq \mathcal{J}_{yy}$) parameterized by the impurity orientation angle $\phi$. Second, the terms strictly proportional to $\mathcal{O}(k_R/k_F)$ couple with $\cos(2k_F R)$, which induces a spatial phase shift $\sin(2k_F R) + \delta \cos(2k_F R) \approx \sin(2k_F R + \delta)$ to the fundamental RKKY oscillations. Because $k_R \ll k_F$ in generic materials, these amplitude corrections and phase shifts are subleading. Throughout the main text, we safely neglect the terms of order $\mathcal{O}(k_R/k_F)$ and $\mathcal{O}(k_R^2/k_F^2)$ to isolate the dominant leading-order spatial beating envelopes and regularized nodal geometries.
\subsection{Derivation of the Dzyaloshinskii-Moriya Interactions}

The antisymmetric DM interactions and the symmetric off-diagonal interactions depend on the cross-products of the scalar, $p$-wave, and Rashba Green's functions. To maintain consistency with the diagonal channels, we retain the unapproximated split momenta $k_\pm = k_0 \pm \kappa$ in the off-diagonal Rashba components.

For the in-plane DM component along the $x$-axis, we must evaluate $iG_0 G_x$. Inserting the expressions for $G_0$ and $G_x$ from Eqs.~(\ref{eq:G0_app}) and~(\ref{eq:Gx_app}), the required product reads
\begin{equation}
i G_0 G_x = i \left( -\frac{im}{4\hbar^2} \right) \left( \frac{m \sin\phi}{4\hbar^2} \frac{k_R}{\kappa k_0} \right) \big[H_+ + H_-\big]\big[k_+ H_+ - k_- H_-\big].
\end{equation}
We expand the momentum difference using the identity $k_+ H_+ - k_- H_- = k_0(H_+ - H_-) + \kappa(H_+ + H_-)$. Combined with the asymptotic Hankel products derived in Eqs. (\ref{eq:H_sum})-(\ref{eq:H_cross}) of the previous subsection, the required bracket evaluates to
\begin{align}
\big[H_+ + H_-\big]\big[k_+ H_+ - k_- H_-\big] &= k_0(H_+^2 - H_-^2) + \kappa(H_+ + H_-)^2 \nonumber \\
&\approx \frac{4}{\pi k_0 R} e^{2ik_0 R} \big[ k_0 \sin(2\kappa R) - i\kappa(1 + \cos(2\kappa R)) \big].
\end{align}
Substituting this back gives the explicit integrand for the cross-product
\begin{equation}
i G_0 G_x \approx \frac{m^2 k_R \sin\phi}{4\pi\hbar^4 \kappa R} e^{2ik_0 R} \left[ -i \frac{\sin(2\kappa R)}{k_0} - \frac{\kappa\big(1 + \cos(2\kappa R)\big)}{k_0^2} \right].
\end{equation}

To obtain the RKKY interaction, we apply the energy integration over the Fermi sea, $\mathcal{J}_{\text{DM},x} = \frac{4J_{\text{imp}}^2}{\pi} \text{Im} \int_0^{k_F} \frac{\hbar^2}{m} k_0 dk_0 (i G_0 G_x)$. The $1/k_0^2$ term is only partially cancelled by the integration measure $k_0 dk_0$, leaving a residual $1/k_0$ factor in the integrand. Applying the rigorous stationary-phase integration rules at the Fermi surface, Eq.(\ref{Eq.68}) and Eq.(\ref{Eq.70}), and factoring out the standard RKKY amplitude $\mathcal{A}(R)$, this directly evaluates to
\begin{equation}
\mathcal{J}_{\text{DM},x}(\bm{R}) \simeq -\mathcal{A}(R) \sin\phi \left[ \frac{k_R}{\kappa} \sin(2\kappa R) \sin(2k_F R) - \frac{k_R}{k_F}\big(1 + \cos(2\kappa R)\big) \cos(2k_F R) \right].
\end{equation}
Following identical steps for $G_y$, which simply substitutes $\sin\phi \to -\cos\phi$ in the overall phase, we obtain
\begin{equation}
\mathcal{J}_{\text{DM},y}(\bm{R}) \simeq \mathcal{A}(R) \cos\phi \left[ \frac{k_R}{\kappa} \sin(2\kappa R) \sin(2k_F R) - \frac{k_R}{k_F}\big(1 + \cos(2\kappa R)\big) \cos(2k_F R) \right].
\end{equation}

For the out-of-plane DM interaction, $\mathcal{J}_{\text{DM},z}$, the cross-product $iG_0 G_z$ is built entirely from components lacking the spatial derivative mapping, yielding
\begin{equation}
i G_0 G_z = i \left( -\frac{im}{4\hbar^2} \right) \left( -\frac{im}{4\hbar^2} \frac{k_M}{\kappa} \right) \big[H_+ + H_-\big]\big[H_+ - H_-\big] = -i \frac{m^2 k_M}{16\hbar^4 \kappa} (H_+^2 - H_-^2).
\end{equation}
This evaluates to $-i \frac{m^2 k_M}{4\pi\hbar^4 \kappa k_0 R} e^{2ik_0 R} \sin(2\kappa R)$. Because the $1/k_0$ factor cancels with the energy integration measure $k_0 dk_0$, no $\mathcal{O}(\kappa/k_F)$ correction is generated. The integral evaluates to the leading-order form
\begin{equation}
\mathcal{J}_{\text{DM},z}(\bm{R}) \simeq -\mathcal{A}(R) \frac{k_M}{\kappa} \sin(2\kappa R) \sin(2k_F R).
\end{equation}
\subsection{Derivation of symmetric off-diagonal interactions}

Finally, the symmetric off-diagonal terms are derived from the cross-products of the Rashba and $p$-wave components. We first consider the in-plane symmetric off-diagonal component $\mathcal{J}_{xy}$, which is governed by the product $G_x G_y$. Inserting the expressions for $G_x$ and $G_y$ from Eqs.~(\ref{eq:Gx_app}) and~(\ref{eq:Gy_app}), the product evaluates to
\begin{equation}
G_x G_y = \left(\frac{m \sin\phi}{4\hbar^2} \frac{k_R}{\kappa k_0}\right)\left(\frac{-m \cos\phi}{4\hbar^2} \frac{k_R}{\kappa k_0}\right) \big[k_+ H_1^{(1)}(k_+ R) - k_- H_1^{(1)}(k_- R)\big]^2.
\end{equation}
Applying the asymptotic substitution $H_1^{(1)}(k_\pm R) \approx -i H_0^{(1)}(k_\pm R) \equiv -i H_\pm$, the momentum bracket squares to $- \big[k_+ H_+ - k_- H_-\big]^2$. Utilizing the asymptotic Hankel products derived in Eqs. (\ref{eq:H_sum})-(\ref{eq:H_cross}) and the identity $\sin\phi\cos\phi = \frac{1}{2}\sin(2\phi)$, this expands to
\begin{equation}
G_x G_y = \frac{m^2 \sin(2\phi)}{32\hbar^4 k_0^2} \frac{k_R^2}{\kappa^2} \big[k_+ H_+ - k_- H_-\big]^2.
\end{equation}
Applying the same integration identities used for the diagonal channels, the energy integration over the Fermi sea yields
\begin{equation}
\mathcal{J}_{xy}(\bm{R}) \simeq \mathcal{A}(R) \frac{k_R^2}{2\kappa^2} \sin(2\phi) \left[ \big(1 - \cos(2\kappa R)\big) \sin(2k_F R) - \frac{\kappa^2}{k_F^2}\big(1 + \cos(2\kappa R)\big)\sin(2k_F R) - \frac{2\kappa}{k_F} \sin(2\kappa R) \cos(2k_F R) \right].
\end{equation}

Next, we evaluate the out-of-plane symmetric off-diagonal terms, governed by the cross-products $G_x G_z$ and $G_y G_z$. For the $xz$ component
\begin{equation}
G_x G_z = - \frac{m^2 \sin\phi}{16\hbar^4 k_0} \frac{k_R k_M}{\kappa^2} \big[k_+ H_+ - k_- H_-\big]\big[H_+ - H_-\big].
\end{equation}
Expanding the brackets and applying the integration mappings yields
\begin{equation}
\mathcal{J}_{xz}(\bm{R}) \simeq -\mathcal{A}(R) \frac{k_R k_M}{\kappa^2} \sin\phi \left[ \big(1 - \cos(2\kappa R)\big) \sin(2k_F R) - \frac{\kappa}{k_F} \sin(2\kappa R) \cos(2k_F R) \right].
\end{equation}

Similarly, for the $yz$ component, mapping $\sin\phi \rightarrow -\cos\phi$ in the spatial derivative gives
\begin{equation}
G_y G_z = \frac{m^2 \cos\phi}{16\hbar^4 k_0} \frac{k_R k_M}{\kappa^2} \big[k_+ H_+ - k_- H_-\big]\big[H_+ - H_-\big].
\end{equation}
Integrating this cross-product over the energy yields
\begin{equation}
\mathcal{J}_{yz}(\bm{R}) \simeq \mathcal{A}(R) \frac{k_R k_M}{\kappa^2} \cos\phi \left[ \big(1 - \cos(2\kappa R)\big) \sin(2k_F R) - \frac{\kappa}{k_F} \sin(2\kappa R) \cos(2k_F R) \right].
\end{equation}

Just as with the diagonal interactions, the inclusion of the momentum splitting systematically generates a Rashba-induced spatial phase shift proportional to $1/k_F$ coupling to $\cos(2k_F R)$ in any channel containing $G_x$ or $G_y$. Neglecting these sub-leading relativistic corrections recovers the regularized, leading-order spatial envelopes utilized throughout the main text.

\section{$f$-WAVE MAGNETS}
\label{app:fwave}

To support the extension to odd-parity magnets, we repeat the derivation of Secs.~\ref{sec:theory}--\ref{sec:results} for the $f$-wave magnet which is defined as,
\begin{align}
M^f_{\boldsymbol{k}}&=\frac{J}{k_F^3}\Big[k_x(k_x^2-3k_y^2)\cos3\beta+k_y(3k_x^2-k_y^2)\sin3\beta\Big] \nonumber \\
&= J\Big(\frac{k}{k_F}\Big)^3\cos3(\theta-\beta),
\label{eq:Mf}
\end{align}
which satisfies $M^f_{-\boldsymbol{k}}=-M^f_{\boldsymbol{k}}$ because $\cos(3\theta+3\pi-3\beta)=-\cos3(\theta-\beta)$. We use Eq.~(\ref{eq:1}) with $M^p_{\boldsymbol{k}}\to M^f_{\boldsymbol{k}}$, and define, as in Sec.\ref{sec:RealSpaceGreenFunction},
\begin{equation}
k_M^f(k,\theta)=\frac{mM^f_{\boldsymbol{k}}}{\hbar^2k}=Q\Big(\frac{k}{k_F}\Big)^{2}\cos3(\theta-\beta),
\qquad
\kappa_f=\sqrt{(k_M^f)^2+k_R^2},
\label{eq:kMf}
\end{equation}
with $Q=mJ/\hbar^2k_F$ and $k_R=m\lambda/\hbar^2$ remaining unchanged. Unlike the $p$-wave case, $k_M^f(k,\theta)$ depends on $k$, so the poles are no longer given by a constant-$\kappa$ factorization.

\subsection{Momentum-space Green's function}
Inverting $[(\mathcal{E}+i\eta)\sigma_0-\hat H(\boldsymbol{k})]$ gives Eqs.~(\ref{eq:11a_new})--(\ref{eq:11d_new}) with $M^p_{\boldsymbol{k}}\to M^f_{\boldsymbol{k}}$ and the common denominator
\begin{equation}
D_f(\boldsymbol{k})=[\mathcal{E}+i\eta-\alpha_{\boldsymbol{k}}]^2-\big[(M^f_{\boldsymbol{k}})^2+\lambda^2k^2\big].
\end{equation}
Crucially, $D_f(\boldsymbol{k})$ remains even in $\boldsymbol{k}$. Consequently, $G_0(\boldsymbol{k})$ is an even function while the vector components $G_{x,y,z}(\boldsymbol{k})$ are odd, ensuring that in real space $G_0(-\boldsymbol{R})=G_0(\boldsymbol{R})$ and $G_{x,y,z}(-\boldsymbol{R})=-G_{x,y,z}(\boldsymbol{R})$. This parity structure depends on the oddness of $M_{\boldsymbol{k}}$. The trace in Eq.~(\ref{eq:H_RKKY}) is therefore unchanged, and Eqs.~(\ref{eq:Jxx})--(\ref{eq:JDMz}) hold for the $f$-wave order.

\subsection{Real-space Green's function}
Writing $\hat H(\boldsymbol{k})=\alpha_{\boldsymbol{k}}\sigma_0+\varepsilon_{\boldsymbol{k}}\hat{\boldsymbol{b}}_{\boldsymbol{k}}\!\cdot\!\boldsymbol{\sigma}$ with $\varepsilon_{\boldsymbol{k}}=\sqrt{\lambda^2k^2+(M^f_{\boldsymbol{k}})^2}$ and $\hat{\boldsymbol{b}}_{\boldsymbol{k}}=(\lambda k_y,-\lambda k_x,-M^f_{\boldsymbol{k}})/\varepsilon_{\boldsymbol{k}}$, the retarded Green's function is
\begin{equation}
\hat G(\boldsymbol{k},\mathcal{E})=\sum_{s=\pm}\frac{\sigma_0+s\,\hat{\boldsymbol{b}}_{\boldsymbol{k}}\!\cdot\!\boldsymbol{\sigma}}{2}\,\frac{1}{\mathcal{E}+i\eta-E_s(\boldsymbol{k})},
\quad E_s(\boldsymbol{k})=\alpha_{\boldsymbol{k}}+s\,\varepsilon_{\boldsymbol{k}} .
\end{equation}
Since $\hat{\boldsymbol{b}}_{-\boldsymbol{k}}=-\hat{\boldsymbol{b}}_{\boldsymbol{k}}$ and $E_s(-\boldsymbol{k})=E_s(\boldsymbol{k})$, each energy contour $E_s(\boldsymbol{k})=\mathcal{E}$ is centrally symmetric and smooth for $k_R>0$. For large impurity separations ($k_FR\gg1$), the spatial Fourier transform is dominated by the stationary-phase point $\boldsymbol{k}_s^*$ where the normal to the Fermi contour is parallel to $\boldsymbol{R}$ (generalizing Eq.~(\ref{eq:A2}) to a non-circular contour),
\begin{equation}
\hat G(\boldsymbol{R},\mathcal{E})\simeq\sum_{s=\pm}\frac{\sigma_0+s\,\hat{\boldsymbol{b}}_{\boldsymbol{k}_s^*}\!\cdot\!\boldsymbol{\sigma}}{2}\;
\Big[-\frac{i}{2\hbar|v_s|}\sqrt{\frac{2\rho_s}{\pi R}}\,e^{i(h_sR-\pi/4)}\Big],
\label{eq:Gband}
\end{equation}
where $v_s$ and $\rho_s$ are the group velocity and radius of curvature at $\boldsymbol{k}_s^*$, respectively. Here, $h_s(\phi)=\max_\theta k_s(\theta)\cos(\theta-\phi)$ is the support function of the contour. For a perfect circle, Eq.~(\ref{eq:Gband}) reproduces Eqs.~(\ref{eq:9_new})--(\ref{eq:12_new}). We define the effective hybridized wavevectors $k_\pm=\bar k\pm\kappa_f$ with $\bar k=\tfrac12(h_-+h_+)$ and $\kappa_f=\tfrac12(h_--h_+)$.

To leading order in the small parameter $\epsilon\equiv Q/k_F$ (which is $\epsilon\ll 1$ in the weak-exchange regime), the contour is described by $k_s(\theta)\simeq k_F[1+\sigma\epsilon\cos3(\theta-\beta)]$ with $\sigma=\pm1$ labeling the spin-polarized branch, yielding:
\begin{equation}
k_M^f(\phi)=Q\cos3(\phi-\beta),\qquad
\kappa_f(\phi)=\sqrt{Q^2\cos^2 3(\phi-\beta)+k_R^2},
\label{eq:kMfphi}
\end{equation}
\begin{equation}
\bar k(\phi)\simeq k_F\Big[1+\epsilon^2\Big(\tfrac52+2\sin^2 3(\phi-\beta)\Big)\Big]+\frac{k_R^2}{2k_F}.
\end{equation}
Eqs.~(\ref{eq:9_new})--(\ref{eq:12_new}) thus maintain their analytical structure with the replacements $k_M\to k_M^f(\phi)$, $\kappa\to\kappa_f(\phi)$, $k_0\to\bar k(\phi)$, and $k_\pm=\bar k\pm\kappa_f$:
\begin{align}
G_0&=-\frac{im}{4\hbar^2}\big[H_0^{(1)}(k_+R)+H_0^{(1)}(k_-R)\big],\\
G_{x}&=\frac{m\sin\phi}{4\hbar^2}\frac{k_R}{\kappa_f}\frac{1}{\bar k}\big[k_+H_1^{(1)}(k_+R)-k_-H_1^{(1)}(k_-R)\big]\\
G_{y}&=\frac{-m\cos\phi}{4\hbar^2}\frac{k_R}{\kappa_f}\frac{1}{\bar k}\big[k_+H_1^{(1)}(k_+R)-k_-H_1^{(1)}(k_-R)\big]\\
G_z&=-\frac{im}{4\hbar^2}\frac{k_M^f}{\kappa_f}\big[H_0^{(1)}(k_+R)-H_0^{(1)}(k_-R)\big],
\label{eq:f-waveGreenfunctionHenkel}
\end{align}
The $f$-wave warping enters strictly through (i) the higher-harmonic angular dependence $\cos(\phi-\beta)\to\cos3(\phi-\beta)$ in $k_M^f$, (ii) the direction-dependent caliper $\bar k(\phi)$, and (iii) relative corrections of order $9\epsilon$ to the amplitudes (from the contour curvature $\rho_s\simeq k_F(1-8\sigma\epsilon\cos3(\phi-\beta))$) and of order $\epsilon^2$ to $\kappa_f$. These corrections are small, and the contour remains strictly convex, provided $Q/k_F\ll 1/9$.

\subsection{Parity-driven decoupling}
Inserting the propagators into Eqs.~(\ref{eq:Jxx})--(\ref{eq:Jzz}), utilizing $A(R)=J_{\rm imp}^2m/(2\pi^2\hbar^2R^2)$, and neglecting $\mathcal{O}(k_R/k_F)$ and $\mathcal{O}(\epsilon)$ sub-leading terms as done in Eqs.~(\ref{eq:J_xx_full})--(\ref{eq:J_zz_full}), we obtain:
\begin{align}
J_{zz}&\simeq-A(R)\Big[\frac{(k_M^f)^2}{\kappa_f^2}+\frac{k_R^2}{\kappa_f^2}\cos(2\kappa_fR)\Big]\sin(2\bar kR),\label{eq:Jzzf}\\
J_{xx}&\simeq-A(R)\Big[\frac{k_R^2\sin^2\phi}{\kappa_f^2}+\frac{(k_M^f)^2+k_R^2\cos^2\phi}{\kappa_f^2}\cos(2\kappa_fR)\Big]\sin(2\bar kR),
\end{align}
while $J_{yy}$ follows from the exchange $\sin\phi \leftrightarrow \cos\phi$. The macroscopic beating $\cos(2\kappa_fR)$ in the Ising channel is therefore weighted solely by $k_R^2/\kappa_f^2$, a structural insulation that holds regardless of unequal band amplitudes. Away from the nodes, this reduces to $J_{zz}\simeq-A(R)\sin(2\bar kR)$ and $J_{xx,yy}\simeq-A(R)\cos(2k_M^fR)\sin(2\bar kR)$, perfectly reproducing the parity-driven role reversal of Eqs.~(\ref{eq:J_xx_simp})--(\ref{eq:J_zz_simp}).

The chiral and symmetric off-diagonal components similarly map onto Eqs.~(\ref{eq:J_xy_sym})--(\ref{eq:J_DMz_reg}) under the same parameter replacements, yielding:
\begin{equation}
J_{{\rm DM},x}\simeq-A(R)\frac{k_R}{\kappa_f}\sin\phi\,\sin(2\kappa_fR)\sin(2\bar kR),\quad
J_{{\rm DM},z}\simeq-A(R)\frac{k_M^f}{\kappa_f}\sin(2\kappa_fR)\sin(2\bar kR).
\end{equation}
Consequently, the massive, nonrelativistic out-of-plane chiral component $J_{{\rm DM},z}$ vanishes along every nodal axis.

\subsection{Nodal $1/R$ window}
The $f$-wave nodal condition $k_M^f=0$ dictates $\cos3(\phi-\beta)=0$, giving the nodal axes $\phi_n=\beta+\pi/6+n\pi/3$. Because $\phi$ and $\phi+\pi$ describe the identical impurity bond axis, the $f$-wave order features \emph{three} distinct nodal axes ($n=0,1,2$), compared to just one in the $p$-wave case. Along these axes, the momentum shift is strictly relativistic ($\kappa_f=k_R$), and the in-plane DM interaction evaluates to,
\begin{equation}
J^{\rm nodal}_{{\rm DM},x}\simeq-A(R)\sin\phi_n\sin(2k_RR)\sin(2\bar kR)
\;\propto\;-\frac{2k_R}{R}\sin\phi_n\sin(2\bar kR),
\end{equation}
valid within the intermediate macroscopic window $1/Q\ll R\lesssim R^*$. Thus, the anomalous dimension-reducing $1/R$ decay and its identifiable slope survive for $f$-wave order. The $f$-wave geometry introduces three physical modifications to the conditions outlined in Sec.~\ref{subsec:decay}:
\begin{enumerate}
\item An angular misalignment $\delta$ from a nodal axis generates a finite shift $k_M^f\simeq\mp 3Q\delta$. Therefore, the effectively 1D-like behavior persists only while $3Q\delta R\lesssim 1$.
\item The nodal gap is exactly $k_R$ only if the stationary point in momentum space remains perfectly on the nodal axis. However, because the azimuthal derivative $\frac{\partial k}{\partial \theta}$ is maximized at the nodes, the normal to the warped Fermi surface tilts. To observe the RKKY interaction along the real-space nodal line $\phi_n$, the stationary momentum must shift off the node by $\delta\theta \simeq \pm \frac{1}{k}\frac{\partial k}{\partial\theta} \sim 3\epsilon$. This geometric shift induces a finite $f$-wave splitting $\kappa_{\rm node} \simeq 9Q^2/2k_F$. The pure Rashba $1/R$ window therefore requires $9Q^2/2k_F \ll k_R$; otherwise, the crossover length $R^*$ is bounded by this curvature-induced splitting rather than the Rashba gap, meaning $R^* \sim 1/(2\max[k_R, \kappa_{\rm node}])$. We note that the specific parameter values $Q/k_F = 0.20$ and $k_R/k_F = 0.05$ used to visually illustrate the macroscopic $p$-wave beating in the main text do not satisfy this strict $f$-wave condition, since $9Q^2/2k_F \approx 0.18 k_F > k_R$. Consequently, observing the extended $1/R$ dimensional reduction in an $f$-wave host demands a comparatively weaker exchange field regime than the $p$-wave case, specifically requiring $Q \ll \frac{1}{3}\sqrt{2k_R k_F}$, which limits this specific $f$-wave crossover to weakly spin-split materials.
\item The fundamental oscillation wavevector is dictated by the direction-dependent caliper $\bar k(\phi)$ rather than the isotropic $k_F$.
\end{enumerate}
Ultimately, the parity-driven decoupling follows intrinsically from the oddness of $M_{\boldsymbol{k}}$ alone, while the nodal $1/R$ window emerges from the vanishing of $k_M$ on symmetry-enforced nodal lines combined with the Rashba gap. Both central phenomena therefore hold universally for any odd-parity magnet within the stated physical conditions.

\end{document}